\documentclass[useAMS,usenatbib]{mn2e}
\usepackage{times}
\usepackage[dvips]{epsfig}
\usepackage[]{natbib}
\usepackage[]{fixltx2e}
\usepackage{booktabs}
\newcommand{\ra}[1]{\renewcommand{\arraystretch}{#1}}
\title[Feeding and Feedback in NGC2110]{Feeding and Feedback in the Inner Kiloparsec of the Active Galaxy NGC2110}
\author[A. Schnorr-M\"uller et al.]
  {Allan Schnorr-M\"uller,$^1$ Thaisa Storchi-Bergmann,$^1$ Neil M. Nagar,$^2$ Andrew Robinson,$^3$ 
  \newauthor Davide Lena,$^3$ Rogemar A. Riffel,$^4$ Guilherme S. Couto$^1$\\
  $^1$Instituto de F\'isica, Universidade Federal do Rio Grande do Sul, 91501-970, Porto Alegre, RS, Brazil\\
  $^2$Astronomy Department, Universidad de Concepci\'on, Casilla 160-C, Concepci\'on, Chile\\
  $^3$Physics Department, Rochester Institute of Technology, Rochester, New York 14623, USA\\
  $^4$Universidade Federal de Santa Maria, Departamento de F\'isica, Centro de Ci\^encias Naturais e Exatas, 97105-900, Santa Maria, RS, Brazil\\}
\date{Released 2013}

\pagerange{\pageref{firstpage}--\pageref{lastpage}} \pubyear{2013}

\begin{document}

\label{firstpage}

\maketitle

\begin{abstract}
We present two-dimensional gaseous kinematics of the inner 1.1\,$\times$\,1.6\,kpc$^2$ of the Seyfert\,2 galaxy NGC\,2110, from optical spectra (5600--7000\,\r{A}) obtained with the GMOS integral field spectrograph on the Gemini South telescope at a spatial resolution of $\approx$\,100\,pc. Gas emission is observed over the whole field-of-view, with complex -- and frequently double -- emission-line profiles. We have identified four components in the emitting gas, according to their velocity dispersion ($\sigma$), which we refer to as: (1) warm gas disk ($\sigma$\,=\,100--220\,km\,s$^{-1}$); (2) cold gas disk ($\sigma$\,=\,60--90\,km\,s$^{-1}$); (3) nuclear component ($\sigma$\,=\,220--600\,km\,s$^{-1}$); and (4) northern cloud ($\sigma$\,=\,60--80\,km\,s$^{-1}$). Both the cold and warm disk components are dominated by rotation and have similar gas densities, but the cold gas disk has lower velocity dispersions and reaches higher rotation velocities. We attribute the warm gas disk to a thick gas layer 
which encompasses the cold disk as observed in some edge-on spiral galaxies. After subtraction of a rotation model from the cold disk velocity field, we observe excess blueshifts of $\approx$\,50\,km\,s$^{-1}$ in the far side of the galaxy (NE)  as well as similar excess redshifts in the near side (SW). These residuals can be interpreted as due to nuclear inflow in the cold gas, with an estimated ionized gas mass inflow rate of $\phi$\,$\approx$\,2.2\,$\times\,10^{-2}$M$_{\odot}$\,yr$^{-1}$. We have also subtracted a rotating model from the warm disk velocity field and found excess blueshifts of $\approx$\,100\,km\,s$^{-1}$ to the SW of the nucleus and excess redshifts of $\approx$\,40\,km\,s$^{-1}$ to the NE, which we attribute to gas disturbed by an interaction with a nuclear spherical outflow. This nuclear outflow is the origin of the nuclear component observed within the inner 300\,pc and it has a mass outflow rate of 0.9\,M$_{\odot}$\,yr$^{-1}$. In a region between 1\arcsec\ and 4\arcsec\ north of the nucleus, which shows strong X-ray and [O\,III]\,$\lambda$5007\r{A} emission, we find a new low $\sigma$ component of ionized gas which we attribute to a high latitude cloud photoionized by the nuclear source. The identification of the 4 distinct kinematic components has clarified the nature of the apparent asymmetry in the rotation curve of the galaxy pointed out in previous studies: it results from the dominance of different components to the south and north of the nucleus. We conclude that a comprehensive two-dimensional coverage of the kinematics and geometry of the nuclear gas around the AGN is necessary to reveal the different processes at play, such as its feeding -- via the cold inflowing gas -- and the feedback, via the warm gas outflows.
\end{abstract}

 
\section{Introduction}

It is widely accepted that the radiation emitted by an active galactic nucleus (AGN) is a result of accretion onto the central supermassive black hole (hereafter SMBH). However, the exact nature of the mechanisms responsible for transferring mass from galactic scales down to nuclear scales to feed  the SMBH is still an open question. Theoretical studies and simulations \citep{shlosman90,emsellem03,knapen05,emsellem06} have shown that non-axisymmetric potentials efficiently promote gas inflow towards the inner regions \citep{englmaier04}. Imaging studies have revealed that structures such as small-scale disks or nuclear bars and associated spiral arms are frequently observed in the inner kiloparsec of active galaxies \citep{erwin99,pogge02,laine03}. While bars can be effective in transporting gas into the inner few hundred parsecs, the fundamental problem of how gas gets from there down to the SMBH has remained unsolved. More recently, \citet{lopes07} found that there is a marked difference in the dust and 
gas content of early-type active and non-active galaxies: while the first always have dusty structures, in the form of spiral and filaments at hundreds of parsecs scales, only 25\% of the non-active ones have such structures. This indicates that a reservoir of gas and dust is a necessary condition for the nuclear activity and also suggests that the dusty structures are tracers of feeding channels to the Active Galactic Nuclei (hereafter AGN).

Previous results by our group in the optical includes the observation of inflow in the central region of NGC\,1097 \citep{fathi06}, NGC\,6951 \citep{thaisa07} and M\,81 \citep{allan11}. In the particular case of M\,81, we could obtain not only the gas kinematics but also the stellar kinematics which was compared to the gaseous kinematics in order to isolate non-circular motions, instead of relying solely on the modeling of the gaseous kinematics as we did for NGC\,1097 and NGC\,6951. With the goal of looking for more cases of inward streaming motions, using all the methodologies we have developed in the previous papers, we began a project to map the gaseous kinematics around additional nearby AGN. After the observation of inflows in three LINER galaxies, we decide to expand our sample to include also more luminous AGN in Seyfert galaxies in order to map the gas flows in a larger a range of AGN luminosities. 

In the current work, we present results obtained from integral field spectroscopic observations of the nuclear region of NGC\,2110, a nearby (30.2\,Mpc, from NED\footnote{NASA/IPAC extragalactic database}, derived assuming a redshift of 0.007789 and H$_{0}$\,=73.0) S0 galaxy harboring a Seyfert 2 nucleus. NGC\,2110 has been the subject of many studies at radio, NIR, optical, UV, and X-ray wavelengths. It is classified as a narrow-line X-ray galaxy \citep{bradt78} because of the strong X-ray emission. Radio observations show extended emission from a well-defined S-shaped radio jet, with an extent of $\approx$\,4\arcsec\ along the North-South direction \citep{ulvestad83,nagar99}, and strong nuclear variability \citep{mundell09}. Optical continuum images show a dusty circumnuclear disk with a spiral pattern and a S-shaped  structure roughly parallel to the radio jet, the northern end of the ``S'' being the site of [O\,III] and soft X-ray emission \citep{evans06}. Spectropolarimetric observations 
reported in \citet{moran07} revealed an extremely broad, double-peaked H$\alpha$ emission line in the polarized flux spectrum, implying the existence of a disk-like hidden broad-line region. Integral field spectroscopy of the inner 10\arcsec\ \citep{delgado02,ferruit04} has shown that the gas kinematics at distances larger than 1\arcsec\ from the nucleus is asymmetric, the northern end of the S-shaped structure being related to the asymmetry. In the inner 1\arcsec, the gas kinematics is complex with at least two components blueshifted relative to the systemic velocity, suggesting the presence of a nuclear outflow \citep{delgado02}. This nuclear outflow was studied by \citet{rosario10} using HST-STIS data. They concluded that the outflow is oriented at a position angle offset by $\approx$40\ensuremath{^\circ} from the PA of the main radio jet, suggesting the nuclear outflow skirts the radio jet instead of being cospatial to it. They also suggested that the S-shaped structure seen in the gas emission is 
ionized by the central AGN and not by shocks as the undisturbed kinematics of 
this region does not support the presence of the fast shock velocities needed to produce its high level of excitation. 
 
The present paper is organized as follows. In Section \ref{Observations} we describe the observations and reductions. In Section \ref{Results} we present the procedures used for the analysis of the data and the subsequent results. In section \ref{Discussion} we discuss our results and present estimates of the mass inflow rate and mass outflow rate and in Section \ref{Conclusion} we present our conclusions.

\section {Observations and Reductions}\label{Observations}

 \begin{figure*}
\includegraphics[scale=0.8]{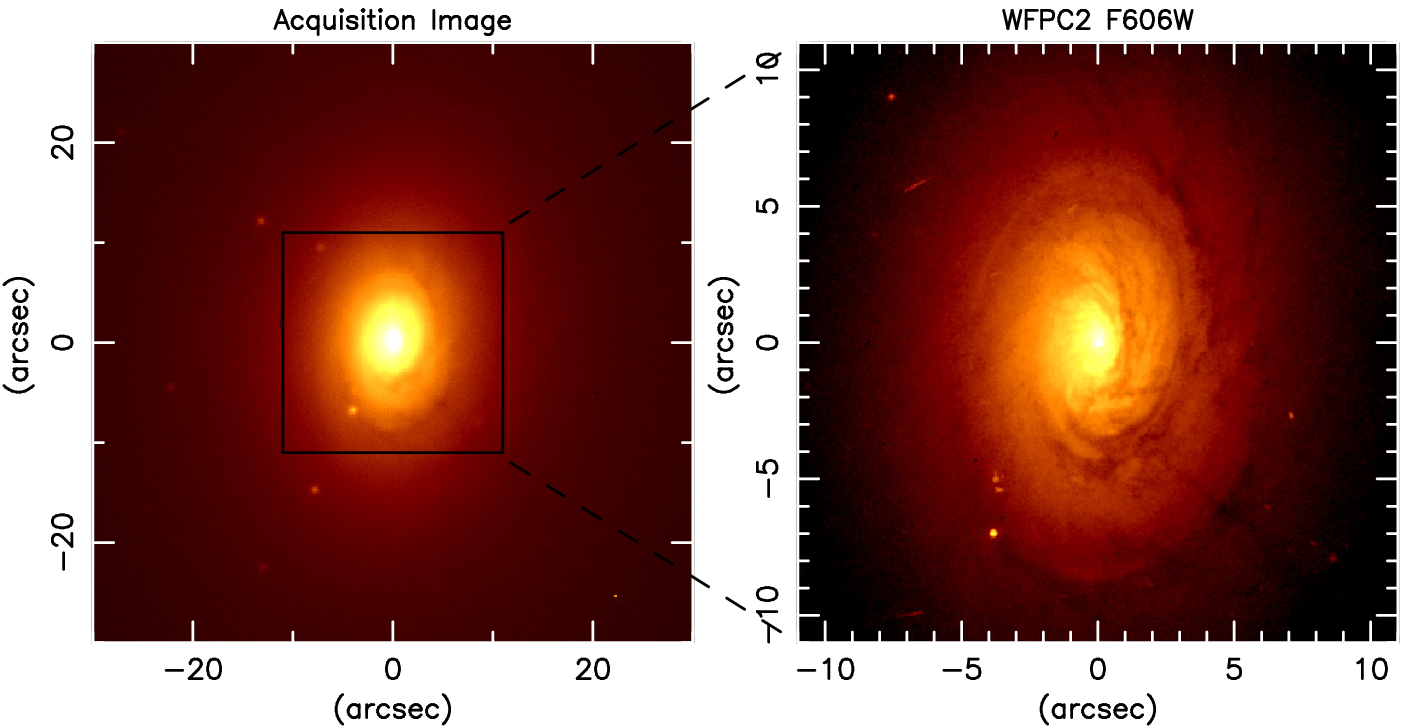}
\includegraphics[scale=0.8]{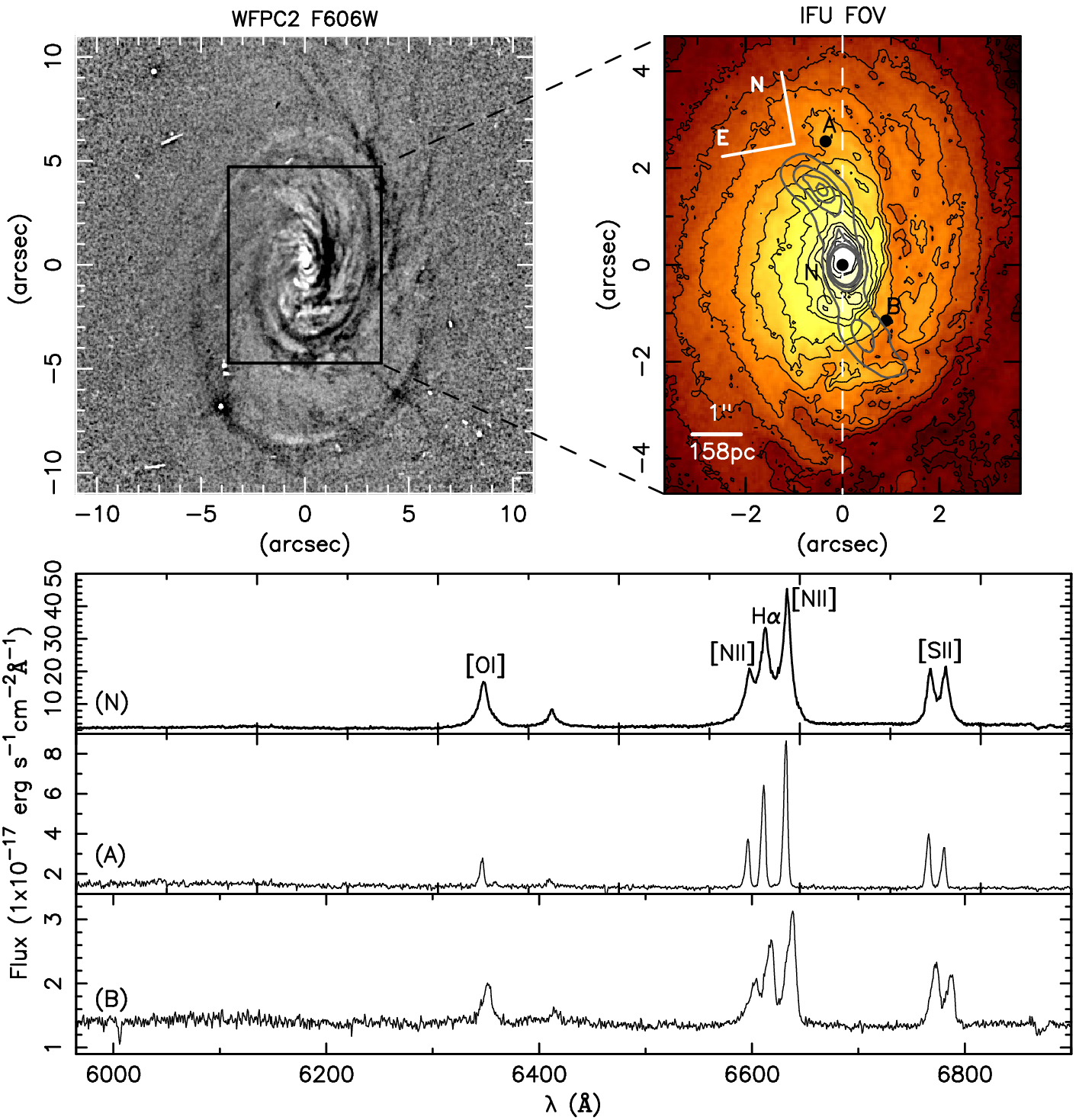}
\caption[Large scale image of NGC\,2110]{Top left: acquisition image. Top right: WFPC2 image of NGC\,2110. Middle left: structure map of the galaxy, the rectangle shows the field of view of the IFU observation. Middle right: zoom in of the WFPC2 image showing the region covered by the IFU observations. Superposed contours trace the radio jet. The dashed white line indicates the position of the major axis of the galaxy (PA\,=\,170\ensuremath{^\circ} \citep{twomass}). Bottom: spectra corresponding to the regions marked as N, A and B in the IFU image.}
\label{fig1}
\end{figure*}

The observations were obtained with the Integral Field Unit of the Gemini Multi Object Spectrograph (GMOS-IFU) at the Gemini South telescope on the night of January 28, 2011 (Gemini project GS-2010B-Q-19). The observations consisted of two adjacent IFU fields (covering 7\,$\times$\,5\,arcsec$^{2}$ each) resulting in a total angular coverage of 7\,$\times$\,10\,arcsec$^{2}$ around the nucleus with a sampling of 0\farcs2. Six exposures of 350 seconds were obtained for each field, slightly shifted and dithered in order to correct for detector defects after combination of the frames. The seeing during the observation was 0\farcs6, as measured from the FWHM of a spatial profile of the calibration standard. This corresponds to a spatial resolution at the galaxy of 95\,pc.

The selected wavelength range was 5600--7000\,\r{A}, in order to cover the H$\alpha$+[N\,II]\,$\lambda\lambda$6548,6583 and [S\,II]\,$\lambda\lambda$6716,6731 emission lines, observed with the grating GMOS R400--G5325 (set to central wavelength of either $\lambda$\,6500\,\r{A} or $\lambda$\,6550\,\r{A}) at a spectral resolution of R\,$\approx$\,2000 with 0.67\,\r{A} per pixel. Absolute flux calibration is expected to be accurate to $\approx\,5\%$, and the wavelength calibration is accurate to the order of 8\,km\,s$^{-1}$.

The data reduction was performed using specific tasks developed for GMOS data in the \textsc{gemini.gmos} package as well as generic tasks in \textsc{iraf}\footnote{\textit{IRAF} is distributed by the National Optical Astronomy Observatories, which are operated by the Association of Universities for Research in Astronomy, Inc., under cooperative agreement with the National Science Foundation.}. The reduction process comprised bias subtraction, flat-fielding, trimming, wavelength calibration, sky subtraction, relative flux calibration, building of the data cubes at a sampling of 0\farcs1$\,\times\,$0\farcs1, and finally the alignment and combination of the 12 data cubes.

\section{Results}\label{Results}

\begin{figure}
\centerline{\includegraphics[scale=0.6]{figure2.eps}}
\caption{Map of the signal-to-noise ratio of the [N\,II]\,$\lambda$6583\r{A} emission line.}
\label{fig2}
\end{figure}

\begin{figure*}
\includegraphics[scale=1.19]{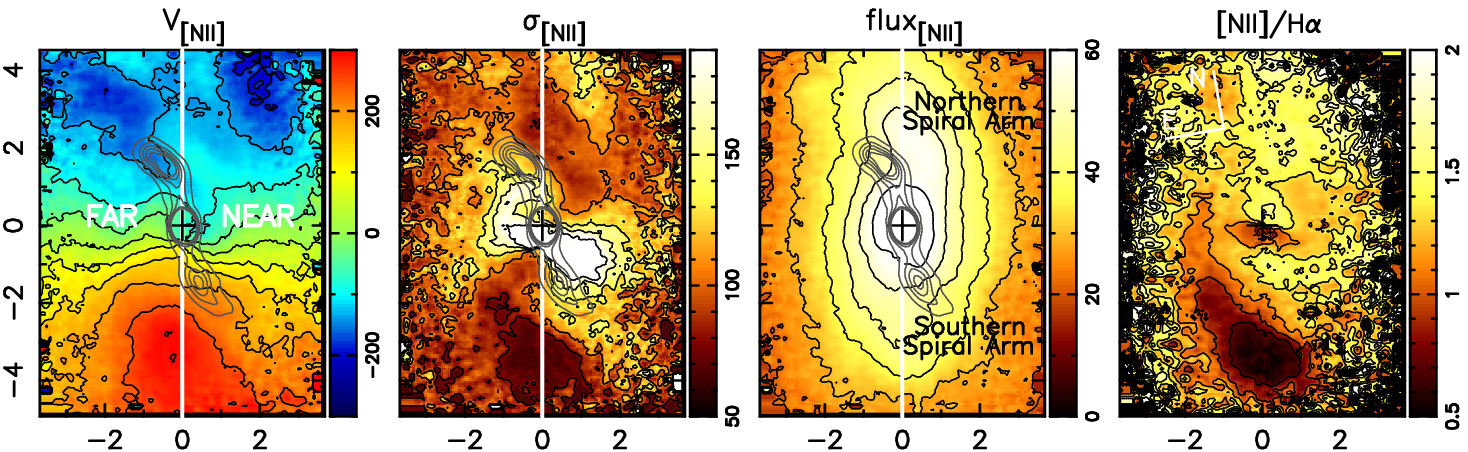}
\caption{Centroid velocity (km\,s$^{-1}$), velocity dispersion (km\,s$^{-1}$), flux distributions (10$^{-17}$\,erg\,cm$^{-2}$\,s$^{-1}$ per pixel), and [N\,II]/H$\alpha$ line ratio resulting from the one component fit.}
\label{fig3}
\end{figure*}

In Fig.\,\ref{fig1} we present in the upper left panel the acquisition image of NGC\,2110 and in the upper right panel an image of the inner 22\arcsec\,$\times\,$22\arcsec\ of the galaxy obtained with the WFPC2 (Wide Field Planetary Camera 2) through the filter F606W aboard the Hubble Space Telescope (HST). In the middle left panel we present a structure map of the WFPC2 HST image of NGC\,2110 (see \citet{lopes07}), where nuclear structures delineated by dark (dusty) spirals can be seen. The rectangle shows the field-of-view (hereafter FOV) covered by the IFU observations. In the middle right panel we present a zoom in of the HST image showing the region covered by our observations. The superposed contours trace the radio jet. Note the presence of two bright spirals roughly (but not exactly) parallel to the radio jet. In the lower panel we present three spectra of the galaxy corresponding to locations marked as A, B and N in the IFU image and extracted within apertures of 0\farcs2\,$\times\,$0\farcs2.

All the spectra are typical of Seyfert 2 galaxies, showing [O\,I]\,$\lambda$$\lambda$\,6300,6363, [N\,II]\,$\lambda$$\lambda$6548,6583, H$\alpha$ and [S\,II]\,$\lambda$$\lambda$6717,6731 emission lines. The spectrum from location A shows very narrow emission lines, while those from the nucleus and from location B show broadened emission lines with blue wings. The nucleus has been assumed to be the location of the peak of the continuum emission.

\subsection{Measurements}

\begin{figure*}
\includegraphics[scale=0.78]{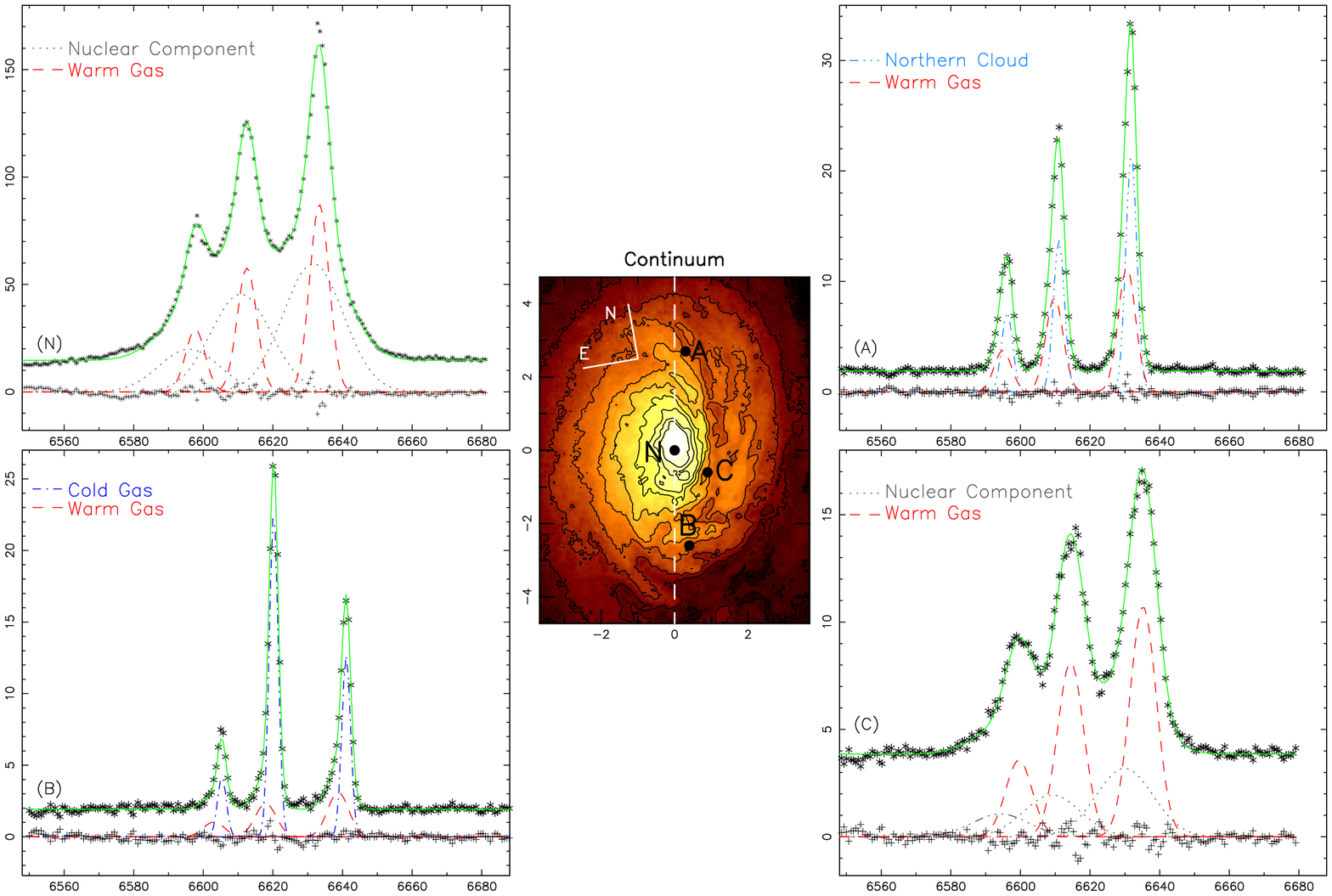}
\caption{Illustration of the two component fit for four different spectra.}
\label{fig4}
\end{figure*}

\begin{figure*}
\includegraphics[scale=1.19]{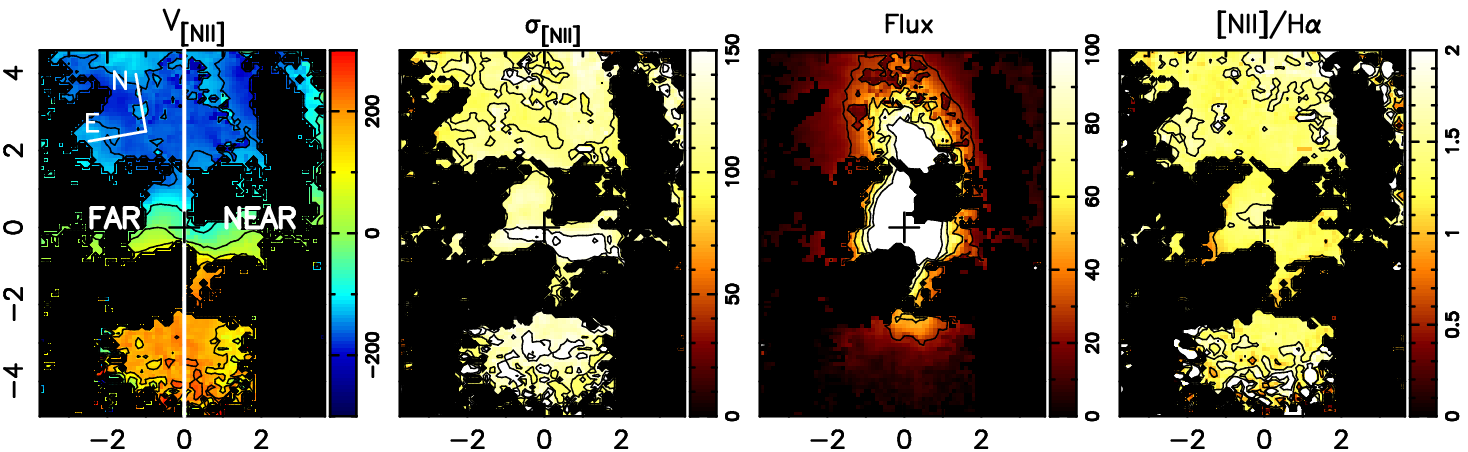}
\includegraphics[scale=1.19]{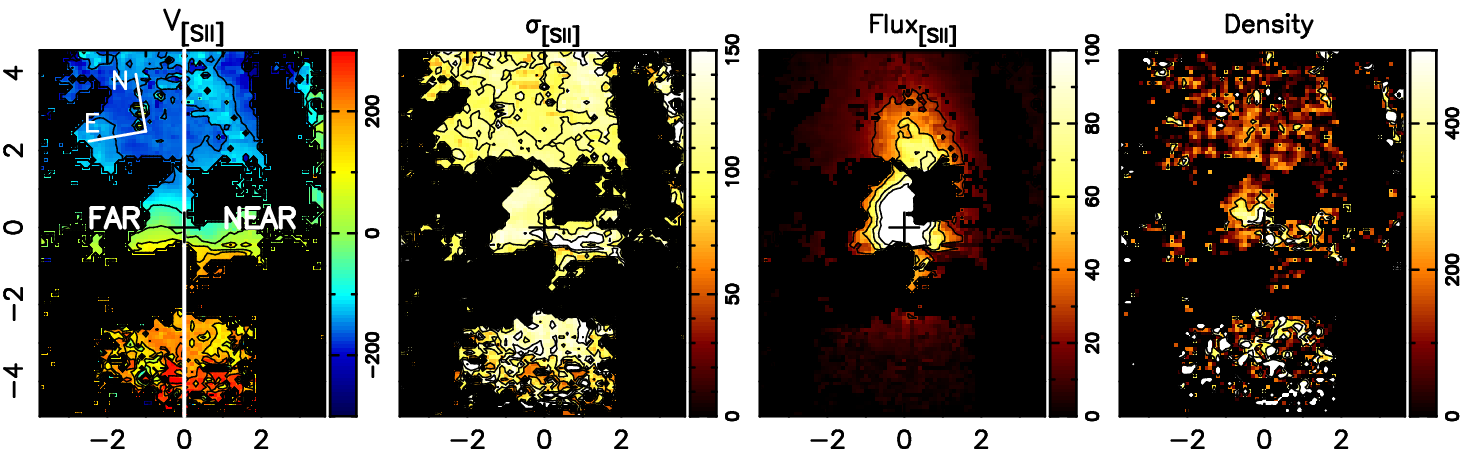}
\includegraphics[scale=1.19]{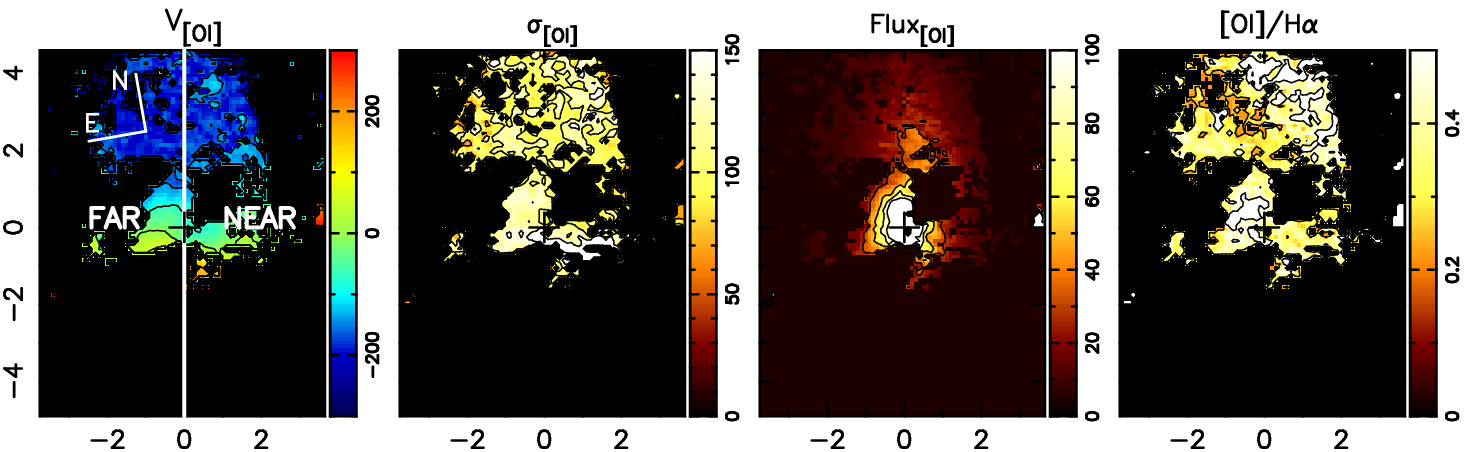}
\caption{Centroid velocity (km\,s$^{-1}$), velocity dispersion (km\,s$^{-1}$), flux distributions (10$^{-17}$\,erg\,cm$^{-2}$\,s$^{-1}$ per pixel), line ratios and density (cm$^{-3}$) of the warm gas disk.}
\label{fig5}
\end{figure*}

\begin{figure*}
\includegraphics[scale=1.19]{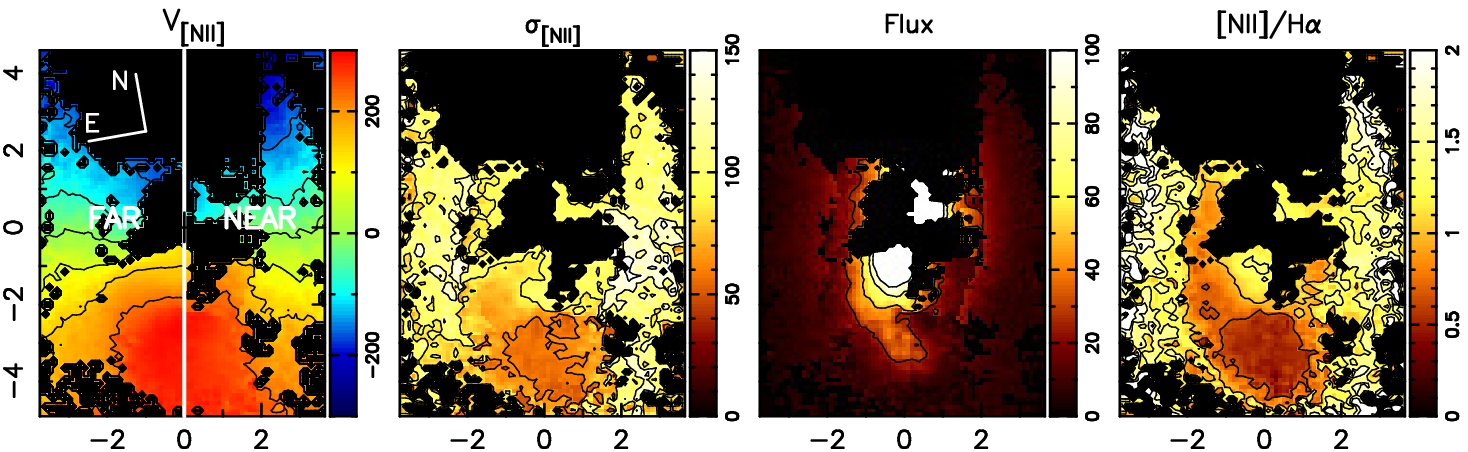}
\includegraphics[scale=1.19]{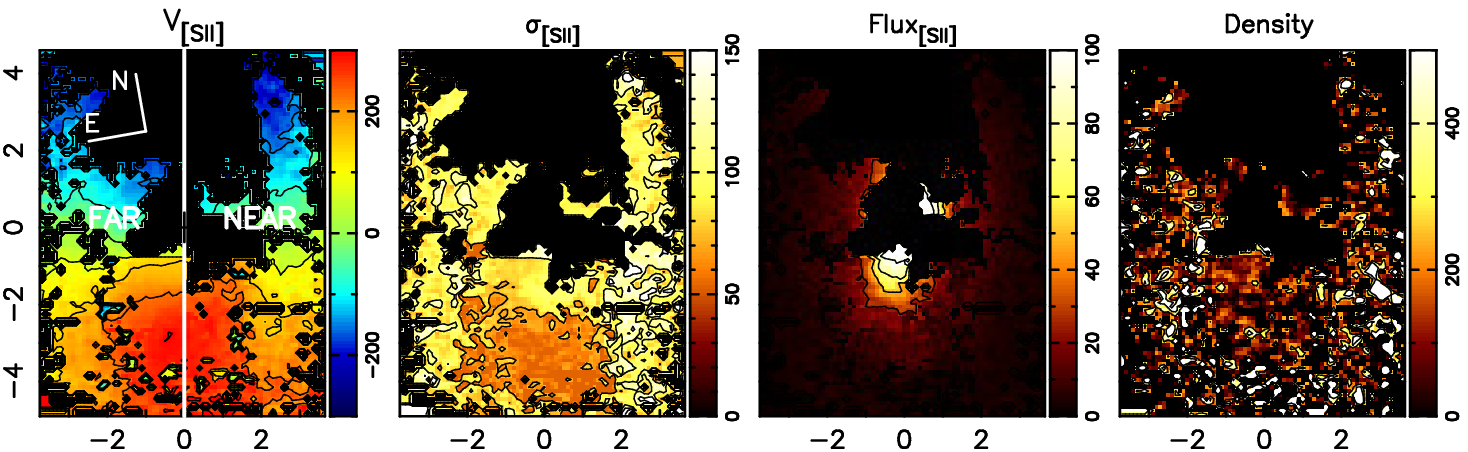}
\includegraphics[scale=1.19]{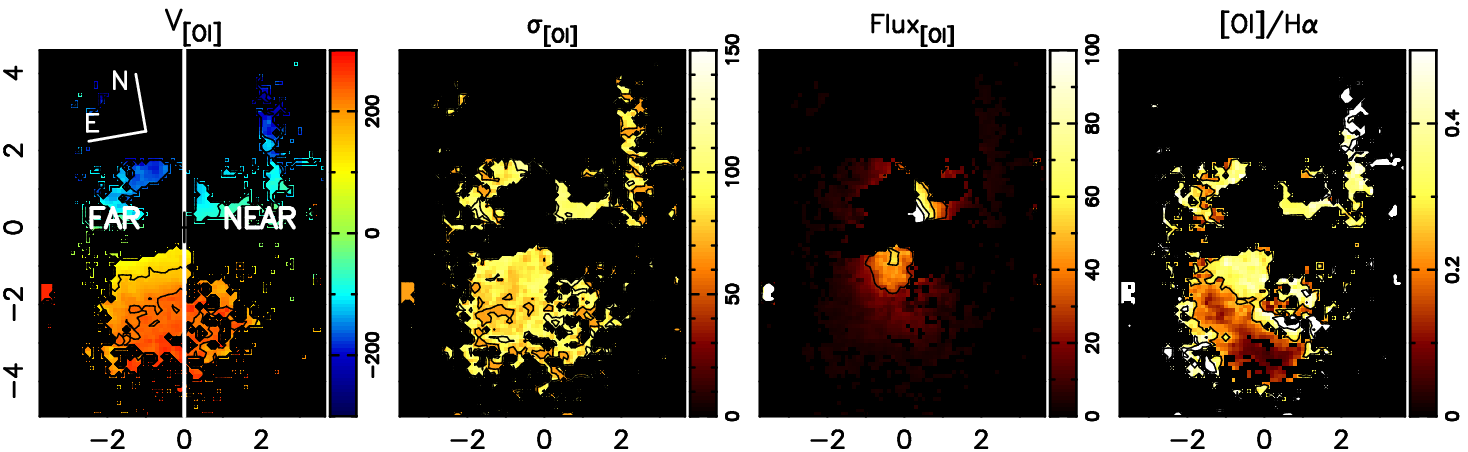}
\caption{Centroid velocity (km\,s$^{-1}$), velocity dispersion (km\,s$^{-1}$), flux distributions (10$^{-17}$\,erg\,cm$^{-2}$\,s$^{-1}$ per pixel), line ratios and density (cm$^{-3}$) of the cold gas disk.}
\label{fig6}
\end{figure*}

\begin{figure*}
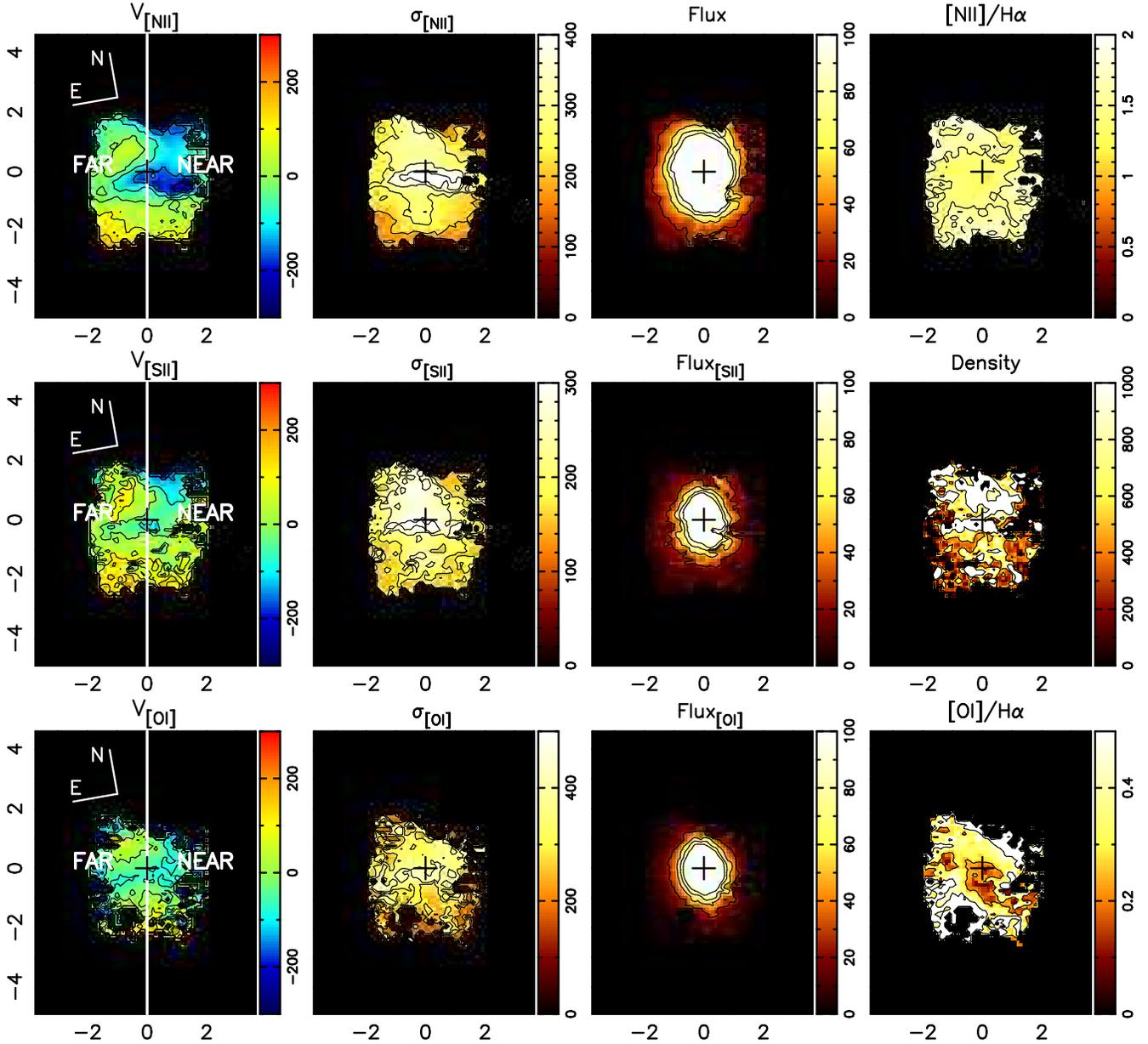

\includegraphics[scale=1.19]{figure7a.eps}
\includegraphics[scale=1.19]{figure7b.eps}
\includegraphics[scale=1.19]{figure7c.eps}
\caption{Centroid velocity (km\,s$^{-1}$), velocity dispersion (km\,s$^{-1}$), flux distributions (10$^{-17}$\,erg\,cm$^{-2}$\,s$^{-1}$ per pixel), line ratios and density (cm$^{-3}$) of the nuclear component.}
\label{fig7}
\end{figure*}

\begin{figure*}
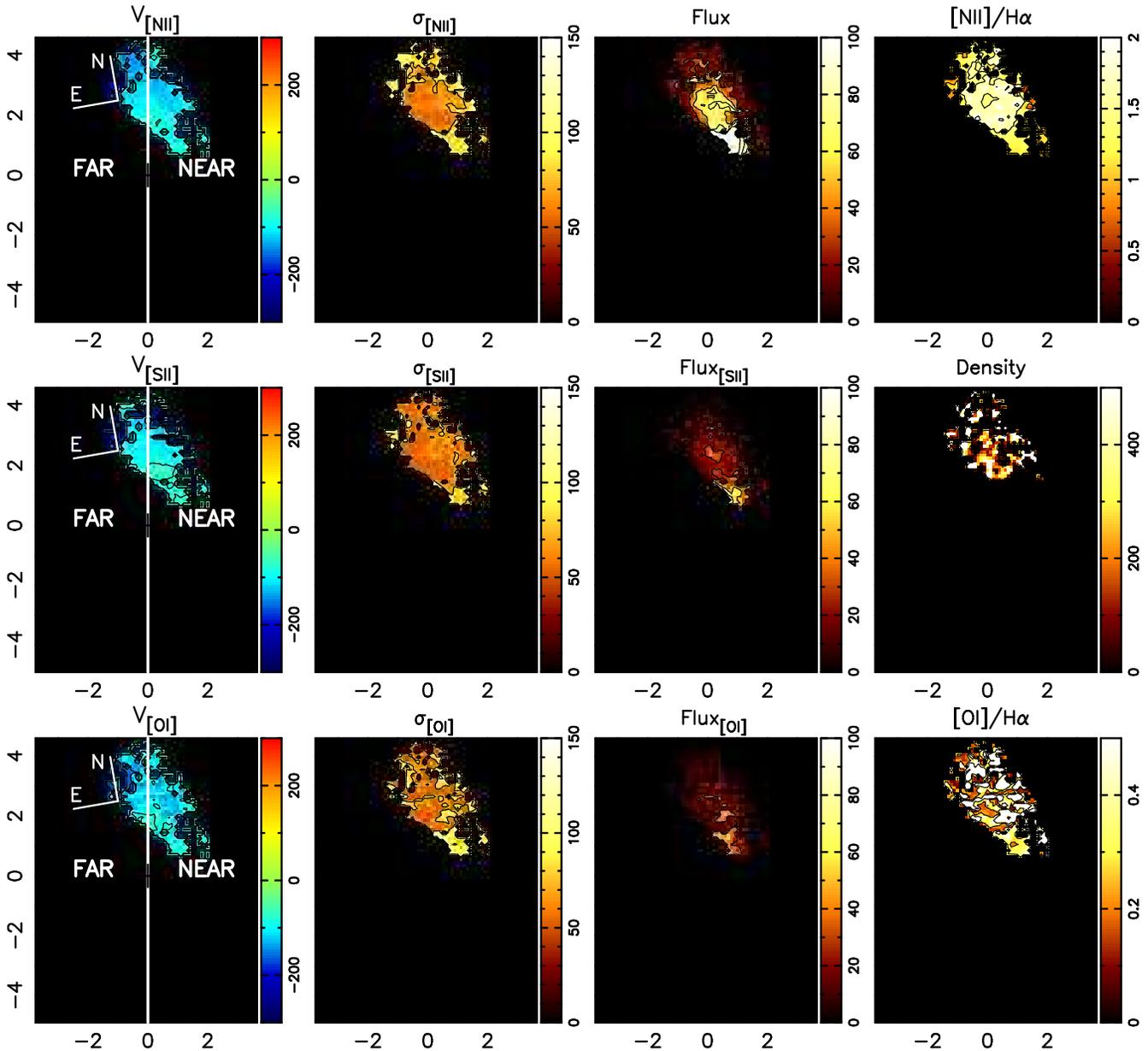

\includegraphics[scale=1.19]{figure8a.eps}
\includegraphics[scale=1.19]{figure8b.eps}
\includegraphics[scale=1.19]{figure8c.eps}
\caption{Centroid velocity (km\,s$^{-1}$), velocity dispersion (km\,s$^{-1}$), flux distributions (10$^{-17}$\,erg\,cm$^{-2}$\,s$^{-1}$ per pixel), line ratios and density (cm$^{-3}$) of the northern cloud.}
\label{fig8}
\end{figure*}

The gaseous centroid velocities, velocity dispersions and the emission-line fluxes were obtained by fitting Gaussians to the [N\,II], H$\alpha$, [O\,I] and [S\,II] emission lines. Since the line profiles are often complex, we performed profile fits with a single Gaussian, two Gaussians, and three Gaussians.
\subsection{The Single Gaussian Fit}

In Fig.\,\ref{fig2} we present a map of the signal-to-noise ratio of the [N\,II]\,$\lambda$6583\r{A} emission line. In Fig.\,\ref{fig3} we present the resulting centroid velocity field, velocity dispersion and flux distribution of the [N\,II] emission line together with the [N\,II/H$\alpha$] ratio obtained from the single Gaussian fit. A systemic velocity of 2229\,km\,s$^{-1}$ (see Section\,\ref{model}) was subtracted from the centroid velocity field. In order to limit the number of free parameters in our fit, we adopted the following physically motivated constraints:

\begin{enumerate}
\item Flux$_{[N\,II]\,\lambda6583}$/Flux$_{[N\,II]\,\lambda6548}=3$;
\item The [N\,II]\,$\lambda$6583 and [N\,II]\,$\lambda$6548 lines have the same centroid velocity and FWHM;
\end{enumerate} 

The centroid velocity map displays a rotation pattern in which the N side of the galaxy is approaching and the S side is receding. Under the assumption that the spiral arms are trailing, it can be concluded that the near side of the galaxy is the W, and the far side is the E. The dust distribution (see the center left panel of Fig.\,\ref{fig1}) also suggests W is the near side of the galaxy. Although the centroid velocity maps display a rotation pattern, clear deviations from simple rotation are present, with the highest distortions observed to the west and north. Note that the rotational velocities observed north of the nucleus, specially in a region near the major axis, are lower than their southern counterpart. This asymmetry was first reported by \citet{wilson85}. More recently, \citet{delgado02} and \citet{ferruit04} showed that the asymmetry is inherent to the gaseous velocity field, as the stellar velocity field is symmetric. 

The velocity dispersion map shows high velocity dispersions ($>$\,200\,km\,s$^{-1}$) in a region extending from 2\arcsec\ SW to 1\farcs5 NE from the nucleus. Velocity dispersions of $\approx$\,180\,km\,s$^{-1}$ are observed in a region extending along the north-south direction located at $\approx$\,2\arcsec\ east of the nucleus. The lowest velocity dispersions ($<$90\,km\,s$^{-1}$) are observed in two spiral shaped regions one to the north and the other to the south of the nucleus. The northern low velocity dispersion region is cospatial to the region of low rotational velocity. 

In the flux distribution we observe two spiral arms approximately parallel to the radio and bending in a similar way. Both spirals are partly cospatial to the low velocity dispersion regions. The [N\,II/H$\alpha$] map shows that the southern low velocity dispersion region (where the southern spiral is located) shows [N\,II/H$\alpha$] $<$\,1. A comparison with centroid velocity map shows that the northern spiral arm is cospatial to a region of low rotational velocities. 
\subsection{The Two Gaussian Fit}

As illustrated by spectra B and N in Fig.\,\ref{fig1}, the emission lines in the inner 1\arcsec\ have strong wings and a single Gaussian fit, although providing a reliable estimate of the centroid velocity, does not provide reliable measurements of the velocity dispersion and flux distribution. \citet{delgado02} and \citet{ferruit04} reported that they successfully fitted the emission line profiles of the inner 1\arcsec\ with two Gaussian components. We thus decided to perform a two-Gaussian fit to our data. As there is no easy way to determine the total extent of the region where a two component fit is necessary, we performed the fit to the entire FOV. We used the continuity of the spectral properties of both components among neighboring pixels as a criteria to decide if a given two component fit was meaningful or not. In addition to the constraints applied to the single Gaussian fit, we assumed that the H$\alpha$ line has the same centroid velocity and FWHM as the [N\,II] lines.  
 
We concluded that the presence of two components is required also beyond the inner 1\arcsec. Roughly, two components were necessary wherever the signal-to-noise ratio was larger than 40 (see Fig.\,\ref{fig2}). The only exception was a region extending from 3\arcsec\ south of the nucleus to the bottom of our FOV, where we observed two components even in regions with signal-to-noise ratio  smaller than 40 (but larger than 20).  Two components were also found for gas extending up to $\approx$\,2\arcsec\ in the east and west directions. In the top half, beyond the inner 1\arcsec, we only found two components along the northern spiral. The kinematics resulting from the two component fit is complex, suggesting that in some locations even more than two components may be necessary in order to reproduce the gas kinematics. Based on the analysis of the centroid velocity and velocity dispersion maps, we concluded we have four components:

\begin{enumerate}
\item A component we called warm gas disk because of its relatively high velocity dispersion, $\sigma$, between 100--220\,km\,s$^{-1}$, and apparent rotation pattern of the centroid velocity;
\item A component we called the cold gas disk because of the low $\sigma$, between 60--90\,km\,s$^{-1}$, and rotation pattern in the centroid velocity field;
\item A component present only within $\approx$\,1\farcs5 from the nucleus, which we call the nuclear component, with $\sigma$ in the range 220--600\,km\,s$^{-1}$;
\item A component only present along the northern spiral arm, with $\sigma$ in the range 60--80\,km\,s$^{-1}$ and centroid velocity close to the systemic velocity, which we called the northern cloud.
\end{enumerate}

We present in Fig.\,\ref{fig4} four characteristic spectra and the respective two component fits. Spectrum N is typical of the inner 0\farcs8: the two components are very broad in this region (200\,km\,s$^{-1}$ for the warm gas disk and 600\,km\,s$^{-1}$ for the nuclear component) and observed in blueshift relative to the systemic velocity. 

Spectrum A is typical of the northern spiral: the emission lines have a broader base and a narrower top relative to a Gaussian profile. One of the components is very narrow ($\approx$\,60\,km\,s$^{-1}$) with a centroid  velocity close to the systemic velocity, attributed to the northern cloud. The other, the warm disk component, is broader ($\approx$\,100\,km\,s$^{-1}$) with a centroid velocity of $\approx$\,--200\,km\,s$^{-1}$. When we fitted this profile using only one component, the fitted Gaussian has essentially the low velocity dispersion of the northern cloud, but the centroid velocity is intermediate between the velocities of the northern cloud and the warm gas disk. 

Spectrum B is typical of the region between 2\arcsec\ and 4\arcsec\ south of the nucleus: the lines are very narrow with a blue wing. The two component fit yields one narrow ($\approx$\,60\,km\,s$^{-1}$) and one relatively broad component ($\approx$\,140\,km\,s$^{-1}$). The narrow component is rotating fast (with velocities of up to $\approx$\,300\,km\,s$^{-1}$) with [N\,II]/H$\alpha$ ratio values typical of H\,II regions ($<$0.5). We attribute this component to cold gas rotating in a thin disk in the plane of the galaxy. The broad component has a centroid  velocity of 200\,km\,s$^{-1}$ and [N\,II]/H$\alpha$ ratio values of the order of 1.2. We attribute this component to the warm gas disk. When fitting only one component to these lines, the narrow component dominates, and the centroid velocities and velocity dispersions we obtain from the one component fit in this region are essentially those of the cold gas disk. We note that although at first glance it may seem that the cold gas disk has stronger line 
emission than the warm gas disk, this is not true. The integrated fluxes from both components are comparable. 

Spectrum C is typical of the high velocity dispersion region SW of the nucleus: a broad component (400\,km\,s$^{-1}$), blueshifted in relation to the systemic velocity, which we attribute to the nuclear component, and a narrower (200\,km\,s$^{-1}$) component, also blueshifted relative to the systemic velocity, which we attribute to the warm gas disk. 

It is worth noting that although we cannot discard the possibility of three, or even four, components being present in a given spectrum, especially in the inner 1\farcs5, a three Gaussian component fit to our data does not give meaningful or stable results. While two of the components are essentially the same as the ones we obtained from the two component fit, the parameters of the third component changed radically between neighbouring pixels and often resulted in non-physical profiles (amplitudes close to zero, extremely high velocities and velocity dispersions). We repeated the fit imposing upper and lower limits to the parameters,  but this did not improve the stability of the parameters of the third Gaussian component.

We also tested the robustness of the two component fits in the inner 1\arcsec. To do this we varied the velocity dispersion of the narrow component from its best fit value and attempted a two component fit. We noticed that if the velocity dispersion of the narrow component differs from its best fit value by more than 10\,km\,s$^{-1}$, the fitted profile does not adequately reproduce either the wings or the peak of the observed profile. Thus we conclude that the two component fit is robust in the inner 1\arcsec.

Having identified the different kinematic components, the next step was to determine which one of these components exists in the regions in which one component fits give better results than two component fits, which we will refer to as the one component regions. Included in the one component regions are the top half of our FOV outside the northern spiral arm and the regions located beyond $\approx$\,2\arcsec\ to the east and west in the bottom half of the field. The centroid velocity field in the one component regions is clearly showing rotation, thus we could only attribute it to either the cold or warm gas disk. We then modeled the velocity fields of the cold and warm gas disks assuming a spherical potential model with pure circular motions \citep{bertola91}, subtracted each model from the centroid velocity field of the one component regions, and used the minimization of the residuals as a criteria to decide which model reproduces it better. We found that in the one component regions we are mapping 
the cold gas disk from the bottom of our FOV to $\approx$\,2\arcsec\ north of the nucleus, and mapping the warm gas disk in the remainder of the FOV. 

\begin{table*}
\scriptsize
\centering
\ra{1.0}
\begin{tabular}{@{}ccccccccccc@{}}\toprule
&\multicolumn{4}{c}{$SN\,>\,80$} & \phantom{abc}& \multicolumn{4}{c}{$80>\,SN\,>60$} &\phantom{abc}\\
\cmidrule{2-5} \cmidrule{7-10}
                   & Cold Disk& Warm Disk& Nuclear Component & Northern Cloud && Cold Disk & Warm Disk & Nuclear Component & Northern Cloud\\ \midrule
Vel$_{[N\,II]}$    & 1        & 2        & 8                 & 2              && 3         & 8         & 13                & 6\\
$\sigma_{[N\,II]}$ & 1        & 3        & 5                 & 2              && 3         & 5         & 10                & 5\\
Flux$_{[N\,II]}$   & 3        & 3        & 1                 & 4              && 4         & 5         & 9                 & 8\\
Flux$_{H\alpha}$   & 1        & 5        & 2                 & 4              && 1         & 6         & 12                & 8\\
Vel$_{[O\,I]}$     & 2        & 2        & 3                 & 5              && 4         & 6         & 7                 & 9\\
$\sigma_{[O\,I]}$  & 3        & 4        & 6                 & 3              && 4         & 8         & 12                & 8\\
flux$_{[O\,I]}$    & 1        & 2        & 3                 & 8              && 7         & 10        & 14                & 12\\
Vel$_{[S\,II]}$    & 2        & 3        & 9                 & 2              && 7         & 6         & 23                & 13\\
$\sigma_{[S\,II]}$ & 2        & 4        & 5                 & 5              && 7         & 11        & 20                & 11\\
Flux$_{[S\,II]}$   & 3        & 3        & 6                 & 5              && 10        & 16        & 21                & 12\\
\hline
&\multicolumn{4}{c}{$60>\,SN\,>40$} & \phantom{abc}& \multicolumn{4}{c}{$SN\,<\,40$} &\phantom{abc}\\
\cmidrule{2-5} \cmidrule{7-10}
                   & Cold Disk& Warm Disk& Nuclear Component & Northern Cloud && Cold Disk & Warm Disk & Nuclear Component & Northern Cloud\\ \midrule
Vel$_{[N\,II]}$    & 3        & 10       & 27                & 10             &&  5        & 20        & -                 & -\\
$\sigma_{[N\,II]}$ & 3        & 5        & 29                & 15             &&  3        & 18        & -                 & -\\
Flux$_{[N\,II]}$   & 5        & 5        & 19                & 12             &&  15       & 20        & -                 & -\\
Flux$_{H\alpha}$   & 2        & 7        & 25                & 13             && 20        & 20        & -                 & -\\
Vel$_{[O\,I]}$     & 15       & 16       & -                 & -              && 22        & -         & -                 & -\\
$\sigma_{[O\,I]}$  & 17       & 19       & -                 & -              && 24        & -         & -                 & -\\
flux$_{[O\,I]}$    & 14       & 20       & -                 & -              && 26        & -         & -                 & -\\
Vel$_{[S\,II]}$    & 3        & 13       & -                 & -              && 11        & 26        & -                 & -\\
$\sigma_{[S\,II]}$ & 5        & 15       & -                 & -              && 11        & 21        & -                 & -\\
Flux$_{[S\,II]}$   & 9        & 18       & -                 & -              && 13        & 23        & -                 & -\\
\bottomrule
\end{tabular}
\caption{ Average uncertainties in velocity (km\,s$^{-1}$), velocity dispersion (km\,s$^{-1}$) and flux ($\%$) for the different components and emission lines. Note that these numbers only reflect uncertainties in the process of fitting Gaussians to the spectra. Absolute uncertainties in these quantities are higher (see Section 2).}
\end{table*}

\subsection{Uncertainties in the Measured Quantities}

To test the robustness of the fits and estimate the uncertainties in the quantities measured from each spectrum in our datacube, we performed Monte Carlo simulations in which Gaussian noise was added to the observed spectrum. For each spaxel, the noise added in each Monte Carlo iteration was randomly drawn from a Gaussian distribution whose dispersion was set to the expected Poissonian noise of that spaxel. One hundred iterations were performed and the estimated uncertainty in each parameter - line center, line width, and total flux in the line - was derived from the $\sigma$ of the parameter distributions yielded by the iterations. Uncertainties vary little among neighboring pixels and become significant only when comparing regions with signal-to-noise ratios that differ by more than 20. Considering this, we divided our FOV into four different regions according to the signal-to-noise ratio of the [N\,II]\,$\lambda$6583\r{A} emission line (see Fig.\,\ref{fig2}) and then calculated the average 
uncertainties inside these regions. The average uncertainties are presented in Table\,1.

\subsection{The Warm Gas Component}

In Fig.\ref{fig5} we present the gaseous kinematics, integrated flux distributions, line ratios and gas densities derived from the [N\,II]\,$\lambda$6583\,\r{A}, H$\alpha$, [O\,I]\,$\lambda$6300\,\r{A} and [S\,II]\,$\lambda$6716\,\r{A} emission lines. 

The gaseous kinematics of the warm gas disk shows a rotation pattern with distortions in the inner 1\arcsec, where there is a strong contribution of the nuclear component. Comparing its velocity field to that of the single Gaussian fit, the warm gas disk centroid velocity field is more symmetrical, with the velocities to the north of the nucleus being only slightly smaller (by $\approx$\,20\,km\,s$^{-1}$) than those to the south. The velocity dispersion varies in the range 110--140\,km\,s$^{-1}$ over most of the FOV, reaching higher values ($\approx$\,210\,km\,s$^{-1}$) along a narrow strip running approximately from 1\farcs5 east of the nucleus to 2\arcsec\ west, cospatial to observed distortions in the centroid velocity field. 

The flux distribution is dominated by emission from the northern part of the FOV. The [N\,II]/H$\alpha$ line ratio has values in the range of 1.2--2.0 in most of the FOV. The [O\,I]/H$\alpha$ line ratio has values in the range 0.3--0.5, with lower values being observed at $\approx$\,2\arcsec\ north of the nucleus. The gas density map, obtained from the [SII]\,$\lambda\lambda$6717/6731 line ratio assuming an electronic temperature of 10000K \citep{osterbrock89}, shows values between 100 and 300\,cm$^{-3}$ over most of the FOV, reaching values larger than 400 in an almost unresolved knot at the nucleus.  

\subsection{The Cold Gas Component}

In Fig.\ref{fig6} we present the gaseous kinematics, integrated flux distributions, line ratios and densities for the [N\,II]\,$\lambda$6583\,\r{A}, H$\alpha$, [O\,I]\,$\lambda$6300\,\r{A} and [S\,II]\,$\lambda$6716\,\r{A} emission lines. 

The gaseous kinematics of the cold gas disk shows a rotation pattern similar to that of the warm gas disk component, but the velocities are higher. In a region extending from 2\arcsec\ south of the nucleus to the borders of the FOV we observe the lowest values of the velocity dispersion, between 60--70\,km\,s$^{-1}$. From 2\arcsec\ to 0\farcs6 south of the nucleus the velocity dispersion increases, reaching 80\,km\,s$^{-1}$. At 1\arcsec\ west and east of the nucleus, the observed velocity dispersion reaches 90\,km\,s$^{-1}$. The regions where the velocity dispersion is higher than 100\,km\,s$^{-1}$, situated at $>$\,2\arcsec\ east and west, are the regions where we observed only one component. 

A conspicuous  feature of the flux distribution and line ratio maps is a southern spiral arm. At the region with lowest velocity dispersion, [N\,II]/H$\alpha$ $<$ 0.5 and [O\,I]/H$\alpha$ $<$ 0.2, values which are typical of [H\,II] regions. Surrounding this region, somewhat higher values are observed along an arm which extends to the east and north, passing to the east of the nucleus. These values are consistent with a combination of photoionization by the central source and young stars. The gas density map does not reveal any particular structure, the densities varying between 100 and 300\,cm$^{-3}$ over most of the FOV. 

\subsection{The Nuclear Component}

Within $\approx$\,2\arcsec\ from the nucleus, the emission lines show a very broad component which we call the nuclear component. In Figs.\ref{fig7} we present the corresponding gaseous kinematics, integrated flux distributions, line ratios and densities for the [N\,II]\,$\lambda$6583\,\r{A}, H$\alpha$, [O\,I]\,$\lambda$6300\,\r{A} and [S\,II]\,$\lambda$6716\,\r{A} emission lines. 

The centroid velocity map shows blueshifted velocities in the near side of the galaxy, from 1\arcsec\ south to 2\arcsec\ north of the nucleus. Velocity dispersions of the order of 400\,km\,s$^{-1}$ are observed east and west of the nucleus, cospatial with the highest velocities. 

In the inner 1\arcsec\ the [N\,II]/H$\alpha$ ratio is $\approx$\,1.5, increasing to $\approx$\,2 at larger radius. The density of the nuclear component reaches very high values, between 800 and 1000\,cm$^{-3}$, in a region extending from 1\arcsec\ south to 1\farcs5 north, decreasing to $\approx$\,300\,cm$^{-3}$ elsewhere. 

\subsection{The Northern Cloud}

We call this component the northern cloud because it shows distinct properties from those of the surrounding gas, as, for example, lower velocity dispersion and centroid velocities. It is cospatial to previously observed strong  [O\,III]$\lambda$5007 and X-ray emission \citep{evans06}.
In Figs.\ref{fig8} we present its gaseous kinematics, integrated flux distributions, line ratios and densities derived from the [N\,II]\,$\lambda$6583\,\r{A}, H$\alpha$, [O\,I]\,$\lambda$6300\,\r{A} and [S\,II]\,$\lambda$6716\,\r{A} emission lines.  

In spite of the small size of the cloud compared to our FOV, we observe a small velocity gradient which is consistent with rotational motion similar to that observed in the disk components, although with smaller velocity values. The velocity dispersion is small, being close to our velocity resolution, in the range 60--80\,km\,s$^{-1}$. 

The high [N\,II]/H$\alpha$ ratios, between 1.4--1.8,  are consistent with photoionization by the central source. The density of the cloud varies between 200 and 500\,cm$^{-3}$. 

\subsection{Channel Maps}
\begin{figure*}
\includegraphics[scale=0.5]{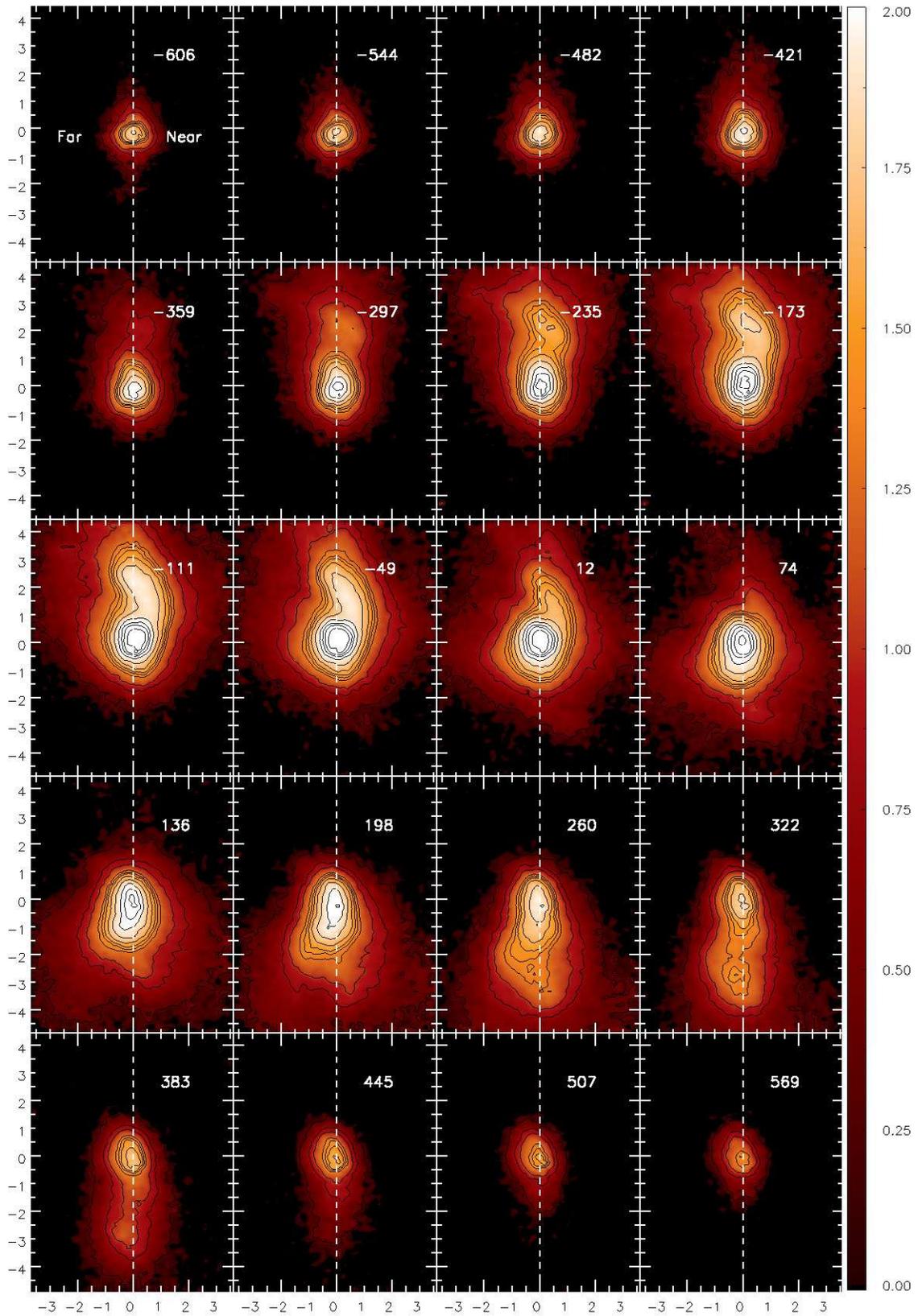}
\caption{Channel maps for the [N\,II] emission line. The dashed white line marks the position of the photometric major axis.}
\label{fig9}
\end{figure*}

We have obtained channel maps along the [N\,II]\,$\lambda$6583\,\r{A} emission line profiles, which are shown in Fig.\,\ref{fig9}. They were built after the subtraction of the H$\alpha$ emission line from the spectra. To test for any contamination due to artifacts resulting from a poor subtraction of the H$\alpha$ line, we obtained channel maps along the [N\,II]\,$\lambda$6548\,\r{A} and compared it with the channel maps for the 6583\,\r{A} line. We found that both channel maps only differ in a region between 1--2\arcsec\ south of the nucleus in the panel corresponding to -606\,km\,s$^{-1}$, where we only detect emission in the 6583\,\r{A} line. Thus we attribute this emission to contamination and we do not consider it in our analysis.

Each panel presents the flux distribution in logarithmic units integrated within the velocity bin centered at the velocity shown in the top-right corner (relative to the systemic velocity of the galaxy). In the panels with velocities between 12\,km\,s$^{-1}$ and -111\,km\,s$^{-1}$, the strong emission from 1\arcsec\ to 4\arcsec\ north of the nucleus, close to the major axis, traces the northern cloud. Those velocities are consistent with what we obtain from the two component fit. The velocity bins between $-235$\,km\,s$^{-1}$ and $-482$\,km\,s$^{-1}$ as well as between 136\,km\,s$^{-1}$ and 445\,km\,s$^{-1}$ are tracing rotation. Note that these panels are roughly symmetric.  The emissions at the highest blueshifts and redshifts, 500--600\,km\,s$^{-1}$, seem to trace the nuclear outflow in the inner 1\arcsec, as they are cospatial to the highest velocity dispersions in Fig.\,\ref{fig7}. We attribute the emissions detected between 1--2\arcsec\ north in the $-544$ and $-606$\,km\,s$^{-1}$ panels and 1--2\arcsec\ south in 
the 507 and 569\,km\,s$^{-1}$ panels to rotating gas.  

\section{Discussion}\label{Discussion}

\begin{figure*}
\includegraphics[scale=0.6]{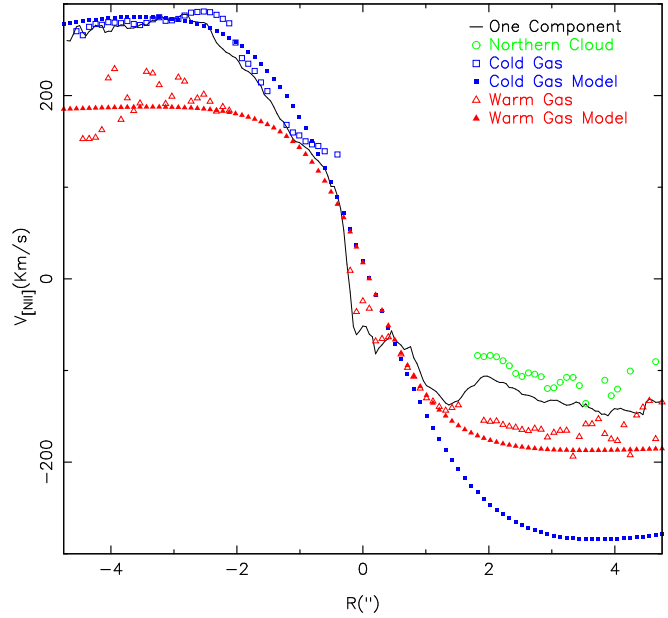}
\caption{Comparison between the velocity curves along the major axis from the one component fit, northern cloud, the cold and warm gas disks and the respective models.}
\label{fig10}
\end{figure*}

\label{gaskinexc}

As already mentioned, previous studies, in both optical and infrared wavelengths, reported an asymmetry in the gaseous velocity field of NGC\,2110 \citep{wilson85,thaisa99,knop01,delgado02,ferruit04}. This asymmetry was observed in the rotation curve along the major axis of the galaxy, which runs along north -- south, with the southern part showing a larger velocity amplitude than the northern part, such that it appeared that the rotation center is displaced to the south relative to the galaxy nucleus. \citet{delgado02} have nevertheless shown that, although the gas velocity field was asymmetric, the stellar rotation was symmetric relative to the galaxy nucleus, with the velocity amplitude being lower than that of the gas. We also observe the asymmetry in the gas velocity field (see Fig.\,\ref{fig3}), as illustrated in Fig.\,\ref{fig10}, where we plot the velocity curve along the major axis for the one component fit, together with those for the different components. It can also be seen in Fig.\,\ref{fig10} that the velocity field of the warm gas disk is practically symmetric relative to the galaxy nucleus (the velocities to the north are only $\approx$\,20\,km\,s$^{-1}$ larger than to the south, but this difference can be credited to uncertainties in the fit of the different components). Although we could not observe emission from the cold gas component along the major axis to the north  (where the emission is completely dominated by the northern cloud and the warm gas), it is possible to observe it in the neighboring regions, and in the next section we show that its velocity field is also symmetric relative to the nucleus. The apparent asymmetry in the one component kinematics is due to the fact that the gas emission is dominated by different components to the north and south of the nucleus: to the south the emission is dominated by the cold gas disk, which shows a larger velocity amplitude than that of the warm disk, while to the north the gas emission is due to both the warm component and the 
northern cloud (see spectrum A in Fig.\,\ref{fig4}); the measured velocities are thus intermediate between those of these two components, $\approx$\,100\,km\,s$^{-1}$ lower than the cold gas disk velocities. Separating the different kinematic components, there is no significant asymmetries in the velocity fields of both the cold and warm disk components, while the northern cloud is apparently at higher disk latitude, showing a smaller velocity than that corresponding to the rotation of the warm disk.

An inspection of Fig.\,\ref{fig10} also illustrates why we cannot separate the cold and warm gas components in the inner 1\arcsec: rotation velocities for both components are similar in these scales. In the region where we observe the outflow the flux seems to be dominated by the warm and nuclear components. Similarly, the northern cloud and the warm gas dominate the emission to the north.

\subsection{Modeling of the Velocity Fields}\label{model}

\begin{figure*}
\includegraphics[scale=1.19]{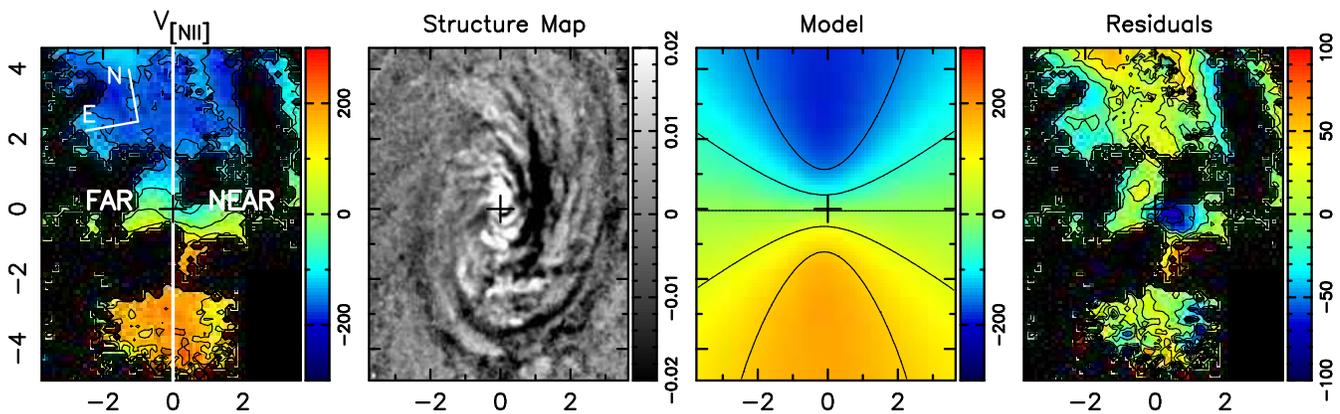}
\caption{Centroid velocity, structure map, modeled velocity field and residuals for the warm gas disk.}
\label{fig11}
\end{figure*}

\begin{figure*}
\includegraphics[scale=1.19]{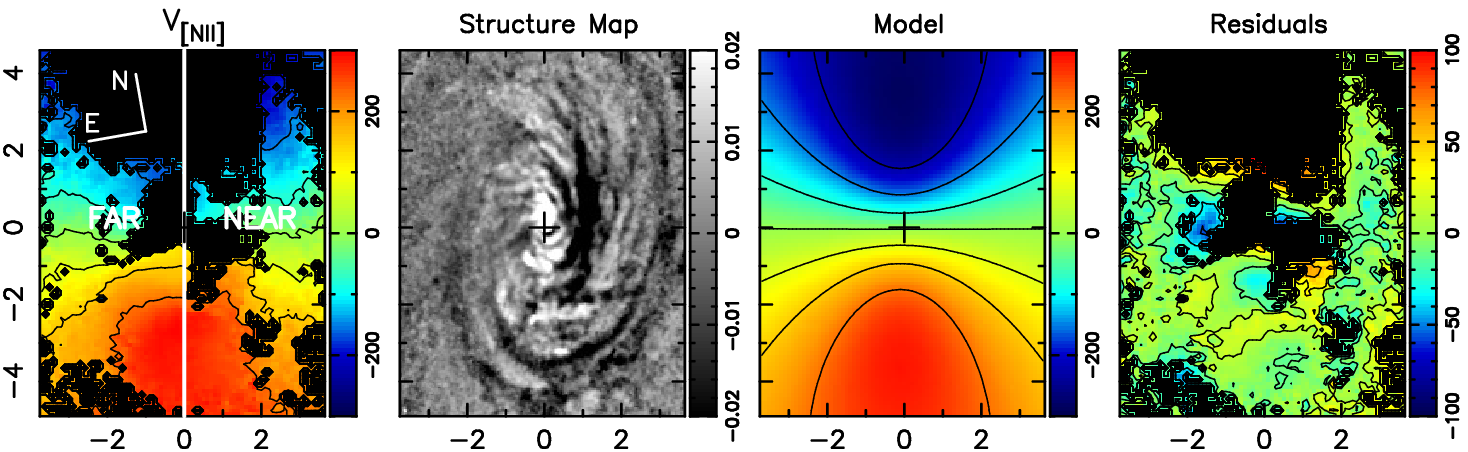}
\caption{Centroid velocity, structure map, modeled velocity field and residuals for the cold gas disk.}
\label{fig12}
\end{figure*}

\begin{figure*}
\includegraphics[scale=1.19]{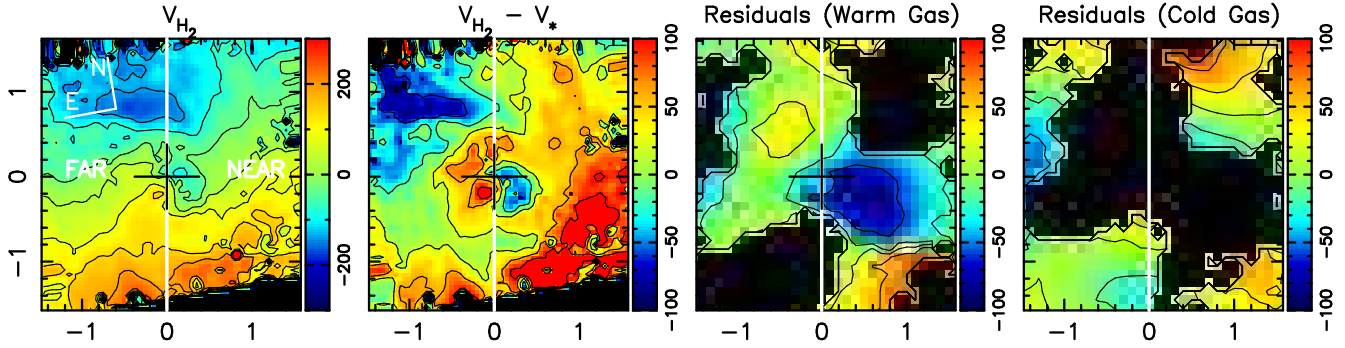}
\caption{H$_2$2.12$\mu$m velocity field from Diniz et al. (in preparation) where we show the velocity field of the H$_2$ emitting gas and its residuals after the subtraction of the stellar velocity field as well as both the cold and warm gas disks after the subtraction of a circular rotation model.}
\label{fig13}
\end{figure*}

In order to isolate non-circular motions in the cold and warm gas disk velocity fields, we modeled both independently assuming a spherical potential with pure circular motions, with the observed radial velocity at a position ($R,\psi$) in the plane of the sky \citep{bertola91} given by:
\footnotesize
\begin{displaymath}
V=V_{s}+\frac{ARcos(\psi-\psi_{0})sin(\theta)cos^{p}\theta}{\{R^{2}[sin^{2}(\psi-\psi_{0})+cos^{2}\theta cos^{2}(\psi-\psi_{0})]+c^{2}cos^{2}\theta \}^{p/2}} 
\end{displaymath}
\normalsize
where $\theta$ is the inclination of the disk (with $\theta$\,=\,0 for a face-on disk), $\psi_{0}$ is the position angle of the line of nodes, $V_{s}$ is the systemic velocity, $R$ is the radius and $A$, $c,$ and $p$ are parameters of the model. We assumed the kinematical center to be cospatial with the peak of the continuum emission. In the case of the warm gas, we fixed the inclination as 42\ensuremath{^\circ} \citep{delgado02}. For the cold gas, however, we left the inclination as a free parameter. The uncertainty in velocity resulting from the uncertainty in the wavelength calibration of 8\,km\,s$^{-1}$, as well as the uncertainties derived from the Monte Carlo simulations (see Sec.\,3.4) were taken into account in the fit.  

For the fit of the cold gas disk centroid velocity field, the resulting parameters $A$, $c$, and $p$ are $1106\,\pm$2\,km\,s$^{-1}$, $2\farcs4\pm0.1$ and $1.4\,\pm0.1$ respectively. The systemic velocity corrected to the heliocentric reference frame is $2309\,\pm$10\,km\,s$^{-1}$ (taking into account errors in the Gaussian fitting process, the velocity field modeling and inaccuracies in the wavelength calibration), the PA of the major axis is 171\ensuremath{^\circ}\,$\pm$1 and the inclination is 39\ensuremath{^\circ}\,$\pm$1. The velocity field of this model is shown in Fig.\,\ref{fig11}. For the warm gas disk, the resulting parameters $A$, $c$, and $p$ are $377\,\pm$4\,km\,s$^{-1}$, $1\farcs3\pm0.1$ and $1.2\,\pm0.1$ respectively. The systemic velocity corrected to the heliocentric reference frame is $2308\,\pm$11\,km\,s$^{-1}$ and the PA of the major axis is 169\ensuremath{^\circ}\,$\pm$1. The velocity field for this model is shown in Fig.\,\ref{fig12}. Considering the good agreement of the systemic velocity obtained from both fits, we adopted $2309\,\pm$10\,km\,s$^{-1}$ as the systemic velocity of the galaxy.

\subsection{Warm Gas Disk Kinematics and Excitation}

The warm gas disk is characterized by velocity dispersions in the range 100--150\,km\,s$^{-1}$. The [N\,II]/H$\alpha$ line ratio ranges between 1.3 and 2 and the [O\,I]/H$\alpha$ between 0.2 and 0.5, which is characteristic of AGNs.

The centroid velocity field of the warm gas disk is dominated by rotation, as evidenced by the small residuals between its centroid velocity field and the best fit circular model shown in Fig.\,\ref{fig11}. Residuals only exceed 30\,km\,s$^{-1}$ in three regions: (1) to the west of the nucleus, (2) in a region $\approx$\,2\arcsec\ south of the nucleus and (3) near the borders of the FOV.  Most of the residuals are of the order of the uncertainties at these locations (see Table.\,1) and thus, although they can be due to non-circular motions, they can also be explained by uncertainties in the fit of the emission-line profiles. For example, the redshifted residuals (between 20--30\,km\,s$^{-1}$) to the north of the nucleus are cospatial to the northern spiral arm, at a location we have found two components, the warm gas disk and the northern cloud. These residuals can thus be attributed to uncertainties in our two component fit due to the small difference in velocity of these two components (about 60\,km\,s$^{-1}$), and not necessarily to the presence of non-circular motions. The largest residuals are observed in blueshifts to the west of the nucleus, reaching 100\,km\,s$^{-1}$ and being cospatial to a region of high velocity dispersion ($\approx$\,200\,km\,s$^{-1}$). They are also cospatial to the highest centroid velocities and velocity dispersions observed in the nuclear component (see Fig.\,\ref{fig7}) and the highest velocity dispersions observed in the one component fit (Fig.\,\ref{fig3}), and we thus suggest that these blueshifted residuals are due to gas in the disk disturbed by a nuclear outflow (see Section\,4.6). 

\subsection{Cold Gas Disk Kinematics and Excitation}\label{coldgasdiscuss}

The cold disk is characterized by velocity dispersions lower than 100\,km\,s$^{-1}$, reaching down to $\approx$\,60\,km\,s$^{-1}$ to the south of the nucleus. The region of low velocity dispersions in the south is delineated in the [N\,II]/H$\alpha$ ratio map by values  $<$\,0.5, which are typical of H\,II regions. We thus argue that the low velocity dispersion corresponds to a region which is now forming stars in an H\,II regions complex. To the north of the complex and to the east of the nucleus, in the far side of the galaxy, there is a region of line-ratio values between 0.5 and 1. Although these values are not typical of H\,II regions, they are lower than those observed in the remaining field, which exceeds 1 everywhere. We thus conclude that in these regions we are observing the superposition of emission from gas photoionized by young stars in the galaxy plane and from gas ionized by the AGN.

The velocity field of the cold gas disk is also dominated by rotation. The residuals between the [N\,II] centroid velocity field and the circular model are shown in Fig.\,\ref{fig12}. They only exceed 30\,km\,s$^{-1}$ at $\approx$\,1\farcs5 northeast and $\approx$\,1\farcs5 southwest of the nucleus, where they are of the order of 50\,km\,s$^{-1}$. Considering that this gas is in the plane and the residuals are in blueshift in the far side of the galaxy and in redshift in the near side, we argue that we are probably observing gas inflowing towards the nucleus of NGC\,2110. Unfortunately, we cannot trace this inflow to smaller scales as the ionized gas emission in the inner 1\arcsec\ is dominated by the warm gas disk and the nuclear component, which do not allow to isolate the cold disk component in this region.

\subsection{Comparison with H$_2$ Molecular Gas Kinematics}

In order to investigate the possibility that we are observing inflowing gas in the cold disk kinematics, we compare our kinematics with that obtained from the H$_2\lambda2.12\mu$m emission line from near-IR integral field spectroscopy using the Gemini NIFS instrument from Diniz et al. (in preparation), who kindly allowed the use of their results here. We present this comparison in Fig.\,\ref{fig13}, where we show the velocity field of the H$_2$ emitting gas and its residuals after the subtraction of the stellar velocity field as well as the residuals from the cold and warm gas disks after subtraction of the respective circular rotation model. A comparison between the H$_2$ and warm gas residual maps shows that the blueshifted residuals west of the nucleus are present in both maps, although restricted to a much smaller region in the H$_2$ map. Another feature of the H$_2$ residuals map is the presence of two spirals, showing blueshifts in the far side of the galaxy (NE) and redshifts in the near side (SW), part of which can also be found in our cold gas residual maps. We thus attribute these residuals to inflowing gas in the plane of the galaxy, in agreement with the interpretation of the H$_2$ residuals by Diniz et al.

\subsection{The Nature of the Cold and Warm Gas Disks}

 Observations of edge-on galaxies show that some of them, mostly late-type spirals with relatively high rates of star formation, have their disks encompassed by a thick layer of ionized gas, usually in the form of thick disks or halos \citep{rand97,rossa03,voigtlander13,rosado13}. Regarding kinematics, these thick disks usually show signs of vertical gradients of rotation velocity or a lag in rotation velocity relative to the disk velocity \citep{heald06a,kamphuis07}, as well as velocity dispersions higher than those characteristic of HII regions \citep{valdez02}. As the characteristics of the warm gas disk are consistent with this, we argue that the warm disk is a thick disk of ionized gas extending to somewhat higher galactic latitudes and encompassing the cold gas disk. 
 
\subsection{The Nuclear Component}

The nuclear component, which is observed within the inner 2\arcsec\ radius is characterized by velocity dispersions in excess of 300\,km\,s$^{-1}$. The FWHM map shows in particular a narrow feature along the galaxy minor axis extending by $\approx$\,1\arcsec\ to the  east and  $\approx$\,2\arcsec\ to the west in which the velocity dispersion reaches 500\,km\,s$^{-1}$. To the south of the nucleus the velocity dispersions are smaller than to the north, $\approx$\,230\,km\,s$^{-1}$, although still higher than the velocity dispersions associated to the cold and warm gas disks. 

The centroid velocity kinematics shows, in the case of [NII] and H$\alpha$, blueshifts larger than $-200$\,km\,s$^{-1}$, which are observed along the high velocity dispersion feature, mainly to the west of the nucleus, with some blueshifts (of $\approx-100$\,km\,s$^{-1}$) also observed to the NW. Some redshifts are observed at $\approx$\,2\arcsec\ to the SE, while in the remainder of the field the centroid velocities are almost zero. 

The blueshifts are observed mostly over the near side of the galaxy and can either be in the disk plane, or at high latitudes or extending from the disk to high latitudes and can thus only be in outflow. Our results are consistent with those found by \citet{delgado02}, who reported the presence of blueshifted gas at the nucleus of NGC\,2110 as well as high velocity dispersions along the minor axis of the galaxy, which they interpreted as due to a nuclear outflow. More recently, \citet{rosario10} -- using HST-STIS optical spectra -- found the [O\,III]$\lambda$5007 emitting gas in the inner 0.2\arcsec\ to be blueshifted by $\approx$\,250\,km\,s$^{-1}$ relative to the systemic velocity and to have velocity dispersions larger than 1200\,km\,s$^{-1}$. Although we do not observe velocity dispersions as high as those reported by \citet{rosario10}, we do find high velocity dispersions at the nucleus, and the blueshifts are similar. Based on their long-slit data and narrow-band imaging, \citet{rosario10} proposed 
that  the nuclear outflow is in a structure they call ``plume", which extends to only 0\farcs5 from the nucleus at PA\,=\,150\ensuremath{^\circ}, making an angle of $\approx$30\ensuremath{^\circ} with the galaxy major axis to the NW. Although we do not have enough spatial resolution to resolve the plume, we do find some blueshifts to the NW at the location of the plume and some redshifts to the SE, as described above, consistent with an orientation of the outflow in the NW-SE direction, as proposed by \citet{rosario10}. On the other hand, the channel maps in Fig.\,\ref{fig9} show that both the highest blueshifted and redshifted velocities are observed in a circular region of radius 2\arcsec\ and the velocity dispersion map for the one component fit (Fig.\,\ref{fig3}) shows that the highest velocity dispersions are observed close to the minor axis, in a ``fan-shaped" structure oriented along NE-SW. The residuals between the warm gas velocity field and the circular rotating model in the inner 1\arcsec\ shown in Fig.\,\ref{fig11}, also show 
blueshifts to the SW and some redshifts (of $\approx$\,30\,km\,s$^{-1}$) to the NE, which could be due to interaction of the nuclear outflow with the warm gas. The residuals between the H$_2$ velocity field and the stellar kinematics shown in Fig.\,\ref{fig13} (Diniz et al., in preparation) also show blueshifts and redshifts approximately along the minor axis of the galaxy, supporting our results. 

One possible interpretation for the nuclear component kinematics and spatial distribution is that of an expanding bubble originating in the galaxy nucleus. An spherically symmetric distribution of clouds ejected from the nucleus will be observed in each spectrum as the combination of gas expanding in different directions, thus resulting in a high velocity dispersion, as observed. The resulting flux distribution would be observed with a round shape, as is the case. In the case of the velocity map, an unobscured spherical outflow would show zero velocity. Dust in the plane of the galaxy, however, would cause obscuration mainly in the redshifted gas emission, which comes from behind the plane. Thus, the observed velocity field would be dominated by blueshifts. A large part of the nuclear component centroid velocity map does show velocities close to zero or slightly blueshifted ($<$\,50\,km\,s$^{-1}$), however regions with larger blueshifts are also observed. An inspection of the structure map in Fig.\,\ref{fig12} shows that the largest blueshifts in the velocity map, mainly observed in the near side of the galaxy, are cospatial to dust lanes, in agreement with obscuration of the redshifted part of the outflow by dust in the plane of the galaxy. The channel maps in Fig.\,\ref{fig9} support our interpretation of the nuclear component as a spherical outflow: both the highest blueshifted and redshifted velocities ($\approx$\,500\,km\,s$^{-1}$) are observed within the inner 300\,pc, with the redshifted channels showing a slightly fainter emission, especially in the near side of the galaxy. In this scenario, we argue that the outflow seen in the warm gas is due to circumnuclear gas in the galaxy plane pushed by the spherical outflow as it moves away from the nucleus.

It is worth pointing out that, although nuclear outflows in nearby Seyfert 2 galaxies such as NGC\,1068 \citep{das06} and NGC\,4151 \citep{thaisa10} have a conical morphology, this is not the case in NGC\,2110.  As discussed above, the nuclear outflow in NGC\,2110 resembles instead spherically symmetric quasar driven outflows. In a recent paper, this was shown to be the case even for type 2 quasars \citep{liu13}.

\subsection{The Northern Cloud}

Previous studies \citep{delgado02,ferruit04} already pointed out that the centroid velocities and velocity dispersions along the northern spiral (which is cospatial with the northern cloud) are lower than those of its surroundings. Also, as we already mentioned, [O\,III] and soft X-ray emission \citep{evans06} are observed along the northern spiral. However, neither the origin of the [O\,III] and soft X-ray emission nor the origins of the lower rotational velocity and velocity dispersion have been explained. \citet{rosario10} argued that a direct interaction of the gaseous disk with the radio jet is unlikely, as they have concluded that the jet escapes the disk towards high latitudes very close to the nucleus (see the cartoon in Fig.\,7 of their paper). They also concluded that the gaseous kinematics along the northern spiral is incompatible with the presence of fast shocks ($>$\,500\,km\,s$^{-1}$) necessary to produce its high level of excitation, concluding that it is photoionized by the central source. 
Taking all this into account, we argue that the [O\,III] emission and soft X-ray emission observed in the so-called ``northern" spiral, originate in fact in the northern cloud. The [N\,II]/H$\alpha$ ratio values of the cloud are similar to those observed in the nuclear outflow, indicating that both regions are photoionized by the central source. We argue that, as in the case of the warm gas disk, the lower rotational velocities observed in the northern cloud compared to the surrounding gas implies that the northern cloud lies at higher latitudes.

\subsection{Estimating the mass outflow rate}

In the case of  a spherically symmetric outflow, we can estimate the mass outflow rate from the ratio of the gas mass and dynamical time, $M_{g}/t_{d}$. From the channel maps in Fig.\,\ref{fig9}, we see that for velocities between 450\,km\,s$^{-1}$ and 600\,km\,s$^{-1}$ the emission is dominated by outflowing gas. We thus adopt an outflow velocity of 525\,km\,s$^{-1}$. Assuming a radius of 316\,pc (2\arcsec) for the nuclear component, we obtain a dynamical time of $t_{d}$\,$\approx$\,6\,$\times$\,10$^{5}$\,yr. The gas mass is given by:

\begin{equation}
{M}_{g}\,=\,N_{e}\,m_{p}\,V\,f
\end{equation}
where $N_{e}$ is the electron density, $m_{p}$ is the mass of the proton, $V$ is the volume and $f$ is the filling factor. The filling factor can be estimated from: 
\begin{equation}
L_{H\alpha}\,\sim\,f\,N_{e}^{2}\,j_{H\alpha}(T)\,V
 \end{equation}
where $j_{H\alpha}(T)$\,=\,3.534$\,\times\,10^{-25}$\,erg\,cm$^{-3}$\,s$^{-1}$ \citep{osterbrock89} and $L_{H\alpha}$ is the H$\alpha$ luminosity emitted by a volume $V$. Substituting equation (2) into equation (1) we have:\begin{equation}
{M}_{g}\,=\,\frac{m_{p}\,L_{H\alpha}}{N_{e}\,j_{H\alpha}(T)}
 \end{equation}

From the H$\alpha$ flux distribution and gas density of the nuclear component we obtain a gas mass of 5.5\,$\times$\,10$^{5}$\,M$_{\odot}$. The mass outflow rate is then $\approx$\,0.9\,M$_{\odot}$\,yr$^{-1}$.

\subsection{Estimating the mass inflow rate}

The cold gas kinematics suggests we are observing inflowing gas to the east and SW of the nucleus (see Fig.\,\ref{fig12} and Section\,\ref{coldgasdiscuss}). We assume that both regions are in the disk plane and the total mass flowing through each of them is the same, a reasonable assumption considering that the densities, residual velocities and flux distributions of both regions are similar. The total gas mass inflow rate will thus be twice that through one region. As the uncertainties in the measurements corresponding to the region with blueshifted residuals are smaller, we have used the properties of this region in the calculation. The mass inflow rate through a cross section of radius $r$ can be obtained from:

\begin{equation}
\dot{M}_{in}\,=\,N_{e}\,v\,\pi\,r^{2}\,m_{p}\,f
\end{equation}
where $N_{e}$ is the electron density, $v$ is the velocity of the inflowing gas , $m_{p}$ is the mass of the proton, $r$ is the cross section radius and $f$ is the filling factor. The filling factor can be estimated from equation (2).
Substituting equation (2) into equation (4) and assuming the volume of the inflowing regions can be approximated by the volume of a cone with a radius $r$ and height $h$, we have:
 
\begin{equation}
\dot{M}_{in}\,=\frac{3\,m_{p}\,v\,L_{H\alpha}}{j_{H\alpha}(T)\,N_{e}\,h}
\end{equation}

We adopt a distance to the nucleus of 1\farcs7. At this distance, the line-of-sight component of the average inflow velocity is 50\,km\,s$^{-1}$ (Fig.\,\ref{fig12}), corresponding to a velocity in the plane of the galaxy of $v$\,=\,79\,km\,s$^{-1}$. The average electron density is 200\,cm$^{-3}$ (see Fig.\,\ref{fig6}). The total H$\alpha$ flux inside the region is 3.0$\,\times\,$10$^{-14}$\,erg\,cm$^{-2}$\,s$^{-1}$. Adopting a distance to the galaxy of 30.2\,Mpc, we obtain $L_{H\alpha}$\,=\,3.2$\,\times\,$10$^{39}$\,erg\,s$^{-1}$. After substituting these quantities into equation (5), we obtain a total inflow rate of $\phi$\,$\approx$\,2.3\,$\times\,10^{-2}$M$_{\odot}$\,yr$^{-1}$. This value is two orders of magnitude lower than our estimate of the mass outflow rate, but we are only observing ionized gas, which is probably only the ``hot skin" of a larger inflow dominated by neutral and molecular gas.

We can now compare the mass inflow and outflow rates in the nuclear region to the mass accretion rate necessary to produce the luminosity of the Seyfert nucleus of NGC\,2110,  calculated as follows:
\[
\dot{m}\,=\,\frac{L_{bol}}{c^{2}\eta}
\]
with $\eta$\,$\approx$\,$0.1$ \citep{frank02} (as usually adopted for Seyfert galaxies). The nuclear luminosity can be estimated from the X-ray luminosity of $L_{X}$\,=\,2.9\,$\times\,$10$^{42}$\,erg\,s$^{-1}$ \citep{pellegrini10}, using the approximation that the bolometric luminosity is $L_{B}$\,$\approx$\,10$L_{X}$. We use these values to derive an accretion rate of $\dot{m}$\,=\,5$\,\times\,$10$^{-3}$\,M$_{\odot}$\,yr$^{-1}$. 

The nuclear accretion rate is 4 times smaller than the mass inflow rate, what shows that much more gas is flowing in than the necessary to feed the AGN (considering in addition that the gas inflow is probably dominated by neutral and molecular gas, as pointed out above).

The nuclear accretion rate is two orders of magnitude smaller than the outflow mass rate, a ratio comparable to those observed for other Seyfert galaxies \citep{rogemar11b}. This implies that most of the outflowing gas does not originate in the AGN, but in the surrounding interstellar medium, which is pushed away from the nucleus by an AGN outflow.

\section{Conclusions}\label{Conclusion}

We have measured the gaseous kinematics in the inner 1.1\,$\times$\,1.6\,kpc$^2$ of the Seyfert\,2 galaxy NGC\,2110, from optical spectra obtained with the GMOS integral field spectrograph on the Gemini South telescope at a spatial resolution of $\approx$\,100\,pc. The main results of this paper are:

\begin{itemize}

 \item The gaseous kinematics is complex and in most of the FOV two components are needed to adequately fit the emission lines. Considering the complete FOV, we identify four distinct kinematic components which we call warm gas disk, cold gas disk, nuclear component and northern cloud;

 \item The cold gas disk has a low $\sigma$ (60--90\,km\,s$^{-1}$) and is rotating with velocities of the order of 300\,km\,s$^{-1}$. A large H\,II region -- characterized by [N\,II]/H$\alpha$ ratios $\le\,0.5$ -- is observed in this component. After subtraction of a rotation model, excess blueshifts to the NE (the far side of the galaxy) and redshifts to the SW (the near side of the galaxy) of $\approx$\,50\,km\,s$^{-1}$ can be interpreted as inflows towards the nucleus;
 
 \item The warm gas disk has 100$\,\le\sigma\le$\,220\,km\,s$^{-1}$ and rotates $\approx$\,100\,km\,s$^{-1}$ slower than the cold disk. We interpret this component as a thick disk of gas extending to higher galactic latitudes and encompassing the cold disk, similar to what is observed in edge-on spiral galaxies. After subtraction of a rotation model, we observe excess blueshifts to the SW of the nucleus and excess redshifts to the NE. We suggest these residuals are due to gas in the disk disturbed by an interaction with the nuclear component. The emission-line ratios are typical of AGN, and we thus conclude that this gas is ionized by the active nucleus;
 
\item The nuclear component has a very high $\sigma$, between 220\,km\,s$^{-1}$ and 600\,km\,s$^{-1}$ and no rotation. We interpret this component as an expanding bubble of gas, a scaled down form of the spherical nuclear outflows characteristic of quasars.;

\item The northern cloud has a much lower velocity dispersion than the surrounding gas, with 60\,$\le\sigma\le$\,80\,km\,s$^{-1}$, shows also a lower rotation velocity than the surrounding gas and its location is the site of X-ray emission previously observed ``along the northern spiral". We interpret this component as due to a gas not in the plane but in a cloud at high latitudes, being ionized by the AGN (as the surrounding gas in the plane);
 
\item In previous observations, this galaxy was found to present an asymmetric rotation curve in the emitting gas. We show that the apparent asymmetry is due to the fact that different locations are dominated (in flux) by distinct kinematic components. When separated, each component, such as the cold and warm disks, have symmetric rotation curves. The presence of the northern cloud and nuclear component also contribute to the apparent asymmetry of the gas velocity field; 

\item From the measured velocities, fluxes and density of the outflowing gas we estimate a mass outflow rate of $\dot{M}_{out}$\,$\approx$\,0.9\,M$_{\odot}$\,yr$^{-1}$. This outflow rate is two orders of magnitude larger than the accretion rate to the AGN, implying that the measured outflow cannot originate in the AGN but instead is due to mass-loading of a nuclear outflow by circumnuclear gas;

\item From the measured gas velocities and fluxes of the inflowing gas -- seen in the cold disk component -- we estimate an ionized gas mass inflow rate of $\phi$\,$\approx$\,2.2\,$\times\,10^{-2}$M$_{\odot}$\,yr$^{-1}$. We argue that this may be only the ``hot skin" of a larger gas flow dominated by neutral and molecular gas. The gas inflow rate is thus at least 4 times larger than the AGN accretion rate, implying that more gas is migrating to the center of the galaxy than the necessary to feed the AGN. 

\end{itemize}
 
With our observations we are thus resolving both the feeding of the AGN -- via the cold inflowing gas-- and its feedback -- via the nuclear outflow in the nuclear component and warm disk -- around the active nucleus of NGC\,2110.

\section*{ACKNOWLEDGMENTS}

We thank Dr. David Rosario for providing comments and suggestions that have improved this paper. This work is based on observations obtained at the Gemini Observatory, which is operated by the Association of Universities for Research in Astronomy, Inc., under a cooperative agreement with the NSF on behalf of the Gemini partnership: the National Science Foundation (United States), the Science and Technology Facilities Council (United Kingdom), the National Research Council (Canada), CONICYT (Chile), the Australian Research Council (Australia), Minist\'erio da Ci\^encia e Tecnologia (Brazil) and south-eastCYT (Argentina). NN acknowledges funding from ALMA-Conicyt 31110016, BASAL PFB-06/2007, Anillo ACT1101 and the FONDAP Center for Astrophysics. This material is based upon work supported in part by the Brazilian institution CNPq. This material is based upon work supported in part by the National Science Foundation under Award No. AST-1108786. 

\bibliographystyle{mn2e.bst}
\bibliography{ngc2110.bib}

\label{lastpage}
\end{document}